\documentclass[preprint,amsmath,amssymb]{revtex4-1}  
\usepackage{graphicx}
\usepackage{slashed}
\usepackage{float}
\usepackage{hyperref}
\usepackage{bm}
\usepackage{url}
\usepackage{accents}

\begin{document}

\bibliographystyle{plain}

\title{General Treatment of Reflection of Spherical Electromagnetic Waves from a Spherical Surface and its Implications for the ANITA Anomalous Polarity Events}

\author{Paramita Dasgupta}
\email{paromita@iitk.ac.in}
\affiliation{Department of Physics, Indian Institute of Technology, Kanpur\\ Kanpur, India- 208016}
\author{Pankaj Jain}
\email{pkjain@iitk.ac.in}
\affiliation{Department of Physics, Indian Institute of Technology, Kanpur\\ Kanpur, India- 208016}
\date{\today}
\begin{abstract}
We develop a general formalism to treat reflection of spherical electromagnetic waves from a spherical surface. Our main
objective is interpretation of radio wave signals produced by cosmic ray interactions with Earth's atmosphere which
are observed by the Antarctica based ANITA detector after reflection off the ice surface. The incident wave is decomposed into
plane waves and each plane wave is reflected off the surface using the standard Fresnel formalism. 
For each plane wave the reflected wave is assumed to be locally a plane wave. 
This is a very reasonable assumption and there are no uncontrolled approximations in our treatment of the reflection phenomenon.
The surface roughness effects are also included by using a simple model. 
We apply our formalism to the radiation produced by the  
balloon-borne HiCal radio-frequency (RF) transmitter. Our final results for the reflected power are found to be
in good agreement with data for all elevation angles.   
We also study the properties of reflected radio pulses in order to study their phase relationship with direct pulses. We find that for some roughness models the pulse shape can be somewhat
distorted and may be misidentified as a direct pulse. However this is a rather small effect
and is unable to provide an explanation for the observed mystery events by ANITA. 
\end{abstract}

\maketitle
\section{Introduction} 

The NASA sponsored balloon-borne ANITA detector ~\cite{Gorham:2010kv,GORHAM200910,PhysRevLett.117.071101,PhysRevD.85.049901} operating in Antarctica is designed 
to detect ultra high energy cosmic rays (UHECR) with energies exceeding 1 EeV ($10^{18})$ eV) \cite{PhysRevLett.105.151101} by collecting the radio pulse generated through the interaction of the primary particle with Earth's
atmosphere ~\cite{Askaryan:1962hbi,1962JPSJS..17C.257A,1965JETP...21..658A}. The radio pulse is detected after reflection from the Antarctic ice
surface. For callibration and measurement of surface reflectivity,
the balloon-borne HiCal radio-frequency (RF) 
transmitter is used. In a recent paper \cite{PhysRevD.98.042004} we developed 
a theoretical
formalism to analyse the process of reflection of such pulses from
the ice surface. The pulse can be considered as a superposition
of spherical waves with a chosen spectral profile. Here we are interested
in determining the mean value of reflection coefficient over the range of
frequencies which are of interest in HiCal observations. Previously, the surface reflectivity was deduced from ANITA-2 \cite{doi:10.1002/2013RS005315} and ANITA-3 observations of the Sun, and also ANITA-3 measurements of HiCal-1 pulses \cite{doi:10.1142/S2251171717400025}.  More details on the instrument can be found in \cite{SCHOORLEMMER201632,2017arXiv171011175G} and \cite{Wang:2017fnm}. 
The formalism we use in this paper  is based on decomposition of a spherical wave in terms of plane waves \cite{Stratton:105763}. This can be represented as
\begin{equation}
{e^{ikr}\over r} = {ik\over 2\pi} \int_0^{2\pi}
\int_0^{{\pi\over 2}-i\infty} e^{ik[x\sin\alpha\cos\beta
+ y \sin\alpha\sin\beta + (z_0-z)\cos\alpha]} \sin\alpha d\alpha d\beta
\label{eq:decomposition}
\end{equation} 
where $\alpha$, $\beta$ are the spherical polar coordinates and this
decomposition is valid 
 in the range $0\le z\le z_0$. 
The reflection properties of each plane wave can be determined by imposing
standard boundary conditions which lead to the 
Fresnel coefficients. The contributions over all such plane waves is added
in order to determine the reflected wave at a given frequency. In 
\cite{PhysRevD.98.042004} it was assumed that the reflected wave corresponding to
each plane wave is also a plane wave. This assumption is not valid for
a curved surface. By comparing our results with the HiCal data it was found to be
in good agreement with observations for elevation angles greater
than $10$ degrees.
The elevation angle is defined as the angle relative to the tangent
at the surface at the point of specular reflection.
 The reason for this agreement is that the dominant contribution
to the reflected wave is obtained from angles $\alpha$, $\beta$ close
to the point of specular reflection. Over such small angles the deviation of reflected
wave from a plane wave may not be very significant. However for small 
elevation angles 
it deviates considerably from observations. Here we develop a more reliable
procedure which only assumes that the reflected wave can be considered locally 
as a plane wave. This is a very reasonable assumption since at any 
point we can define a tangent plane to the wave front which provides a good approximation to the wave front in a small
neighbourhood of that point.

\section{Reflection and Transmission on a Flat Surface} 
\label{eq:flat}
\begin{figure}[H]
\begin{centering}
\includegraphics[scale=0.8]{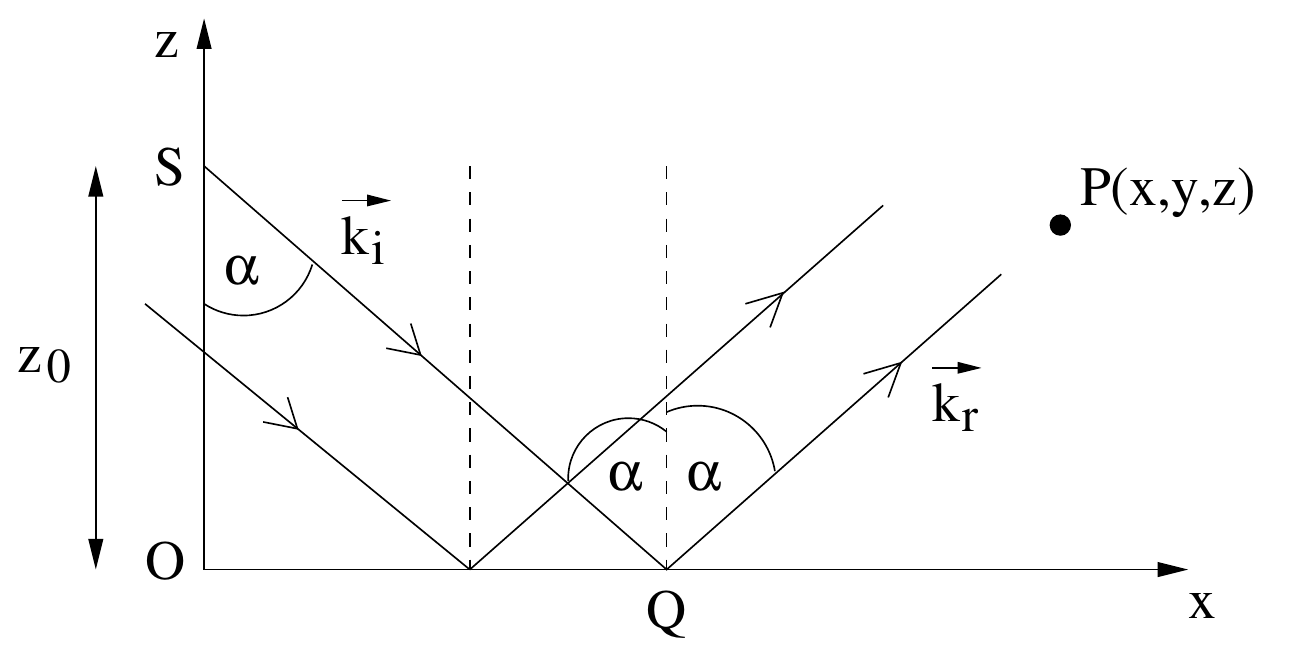}
\par\end{centering}
{\caption{\it Reflection from a flat surface: the dipole source is 
located at $S (0,0,z_{0})$ and the detector is located at $P$ whose 
position vector is $\vec r=(x,y,z)$ with respect to the origin $O$. 
A plane wave for which $\vec k_i$ makes an angle $\pi-\alpha$
 with the $z$ axis is shown. The point $Q$ is the position on the
flat surface where the wave vector $\vec k_i$ directed from $S$ strikes
the surface. 
}\label{fig:geometry1}}
\end{figure}

{}
We start by reviewing the formalism for the case of a flat surface since
the resulting formulas would be used in the case of spherical surface.
The basic geometry for this case 
is shown in Fig. \ref{fig:geometry1}. 
Here the source $S$ lies at $(0,0,z_0)$.  
Let $\vec{E}_{q}$ denote the electric field vector for a 
particular incident plane wave. 
The subscript $q=i,r,t$ designates the incident, reflected and transmitted 
waves respectively.
The corresponding magnetic field is denoted by $\vec{H}_q$. 
Here we are primarily interested in the H-pol which corresponds to the component of the 
electric field perpendicular to the plane of incidence.
The complete Hertz potential for the direct wave is given by 
\begin{equation}
\vec{\Pi}_{dir}=\frac{e^{ikr}}{4\epsilon\pi r} \hat{y}
\label{eq:HertzDirect}
\end{equation}
We decompose this into plane waves and
for a given plane wave the Hertz potential can be written as 
\cite{PhysRevD.98.042004}
\begin{equation}
\vec{\Pi}_{inc}=\frac{ik}{8\epsilon\pi^{2}}
\tilde\Pi
\hat{y}
\end{equation}
where
\begin{equation}
\tilde\Pi = 
e^{ikz_{0}\cos{\alpha}}
e^{ik(x\sin\alpha\cos\beta+y\sin\alpha\sin\beta-z\cos\alpha)}
\end{equation}
The electric and magnetic fields can be computed by using the formula, 
\begin{eqnarray}
\vec E &=& \vec \nabla (\vec \nabla\cdot \vec \Pi) + k^2\vec \Pi\nonumber\\ 
\vec{H} &=& \frac{k^{2}}{i\omega\mu}(\vec\nabla\times\vec{\Pi})
\end{eqnarray}
where, $\omega$ is the angular frequency of radiation and $\mu$ is the permeability of medium.

For an incident wave vector given by
\begin{equation}
 \vec k_i= k[\sin\alpha\cos\beta\hat x+ \sin\alpha\sin\beta\hat y
-\cos\alpha \hat z]
\label{eq:kinc}
\end{equation}
the unit normal $\hat \eta$ to the plane of incidence 
is given by 
\begin{equation}
\hat \eta = l\hat{x}+m\hat{y}+n\hat{z}\,. 
\end{equation}
We have $\vec k_i\perp\hat \eta$ and $\hat z\perp\hat \eta$. 
This leads to
$\hat \eta = (-\sin\beta\hat{x}+\cos\beta\hat{y})$. 
The incident electric and magnetic fields in the far zone, $r>>\lambda$,
 are given by  
\cite{PhysRevD.98.042004}
\begin{eqnarray}
\vec{E}_{i}&=& \frac{ik^{3}}{8\epsilon\pi^{2}}\tilde\Pi
\left[-\sin^{2}\alpha\cos\beta\sin\beta\hat{x}+(1-\sin^{2}\alpha\sin^{2}\beta)\hat{y}+(\sin\alpha\sin\beta\cos\alpha)\hat{z}\right]\nonumber\\
\vec{H}_{i}&=& \frac{ik^{2}\omega}{8\pi^{2}}\tilde\Pi
\left[\cos\alpha\hat{x}+(\cos\beta\sin\alpha)\hat{z}\right]
\label{eq:incident}
\end{eqnarray}
We split these into components perpendicular and parallel
to the plane of incidence, i.e., 
\begin{equation}
 \vec{E}_{q}= \vec{E}^{s}_{q}+\vec{E}^{p}_q \nonumber
\end{equation}
\begin{equation} \label{eq:electric_1}
 \vec{H}_q= \vec{H}^{s}_q+\vec{H}^{p}_q
\end{equation}
 For the electric (magnetic) field, $\perp$ and $\parallel$ components are 
denoted 
by the superscripts
$s$ ($p$) and $p$ ($s$), respectively. 

The $s$ and $p$ components of $\vec E_i$ can be
expressed as:
\begin{equation}
 \vec{E}^{s}_{i}=  \hat \eta [\vec{E}_{i}\cdot \hat \eta]=\frac{ik^{3}}{8\epsilon\pi^{2}}\tilde\Pi
\, (-cos\beta\sin\beta\hat{x}+\cos^{2}\beta\hat{y})
\end{equation}
\begin{equation}\ 
\vec{E}^{p}_{i}=  \frac{ik^{3}}
{8\epsilon\pi^{2}}\tilde \Pi
(\cos^{2}\alpha\cos\beta\sin\beta\hat{x}+\cos^{2}\alpha\sin^{2}\beta\hat{y}+\sin\alpha\cos\alpha\sin\beta\hat{z})
\label{eq:electric_2} \,.
\end{equation}
Similarly, for the magnetic field  
\begin{equation}
\vec{H}^{p}_{i}=[\vec{H}_{i}\cdot \hat \eta] 
\hat\eta=\frac{ik^{2}\omega}{8\pi^{2}}\tilde\Pi
(\cos\alpha\sin^{2}\beta\hat{x}-\cos\alpha\cos\beta\sin\beta\, \hat{y}) 
\end{equation}
\begin{equation}
\vec{H}^{s}_{i}=   
\frac{ik^{2}\omega}{8\pi^{2}}\tilde\Pi
(\cos\alpha\cos^{2}\beta\hat{x}+\cos\alpha\cos\beta\sin\beta\hat{y}+\sin\alpha\cos\beta\hat{z}) \,.
\end{equation}
The components of the reflected wave are 
given by
\begin{equation}
 \vec{E}^{s}_{r}
  =f^{s}_{r}\frac{ik^{3}}{8\epsilon\pi^{2}}\tilde \Pi_{r}
(-\cos\beta\sin\beta\hat{x}+\cos^{2}\beta\hat{y})
\label{eq:erefs}
\end{equation}
\begin{equation}
\vec{E}^{p}_{r}= f^{p}_{r}\frac{ik^{3}}{8\epsilon\pi^{2}}\tilde\Pi_{r}
(-\cos^{2}\alpha\cos\beta\sin\beta\hat{x}-\cos^{2}\alpha\sin^{2}\beta\hat{y}+\sin\alpha\cos\alpha\sin\beta\hat{z})
\label{eq:erefp}
\end{equation}
where
\begin{equation}
\tilde\Pi_{r} = 
e^{ikz_{0}\cos{\alpha}}
e^{ik(x\sin\alpha\cos\beta+y\sin\alpha\sin\beta+z\cos\alpha)} \,.
\end{equation}
Similarly,  
\begin{equation}
\vec{H}^{p}_{r}
=f^{p}_{r}\frac{ik^{2}\omega}{8\pi^{2}}\tilde\Pi_{r}
(\cos\alpha\sin^{2}\beta\hat{x}-\cos\alpha\cos\beta\sin\beta\hat{y})
\end{equation}
and
\begin{equation}
\vec{H}^{s}_{r}= f^{s}_{r}\frac{ik^{2}\omega}{8\pi^{2}}\tilde\Pi_{r}
(-\cos\alpha\cos^{2}\beta\hat{x}-\cos\alpha\cos\beta\sin\beta\hat{y}+\sin\alpha\cos\beta\hat{z})\,.
\end{equation}
Here $f^{s}_{r}$  and $f^{p}_{r}$ are the reflection coefficients
which are determined by boundary conditions.

The corresponding transmitted fields $\vec{E}^{s}_{t}$, $\vec{E}^{p}_{t}$, $\vec{H}^{s}_{t}$ and $\vec{H}^{p}_{t}$ 
are given by
\begin{equation} \label{eq:eflattrans}
\vec{E}^{s}_{t}= f^{s}_{t}\frac{ik_{1}^{3}}{8\epsilon_{1}\pi^{2}}
(-\cos\beta_{t}\sin\beta_{t}\hat{x}+\cos^{2}\beta_{t}\hat{y})\tilde\Pi_t
\end{equation}
\begin{equation} \label{eq:eflattranp}
\vec{E}^{p}_{t}=
f^{p}_{t}\frac{ik_{1}^{3}}{8\epsilon_{1}\pi^{2}}
(\cos^{2}\alpha_{t}\cos\beta_{t}\sin\beta_{t}\hat{x}+\cos^{2}\alpha_{t}\sin^{2}\beta_{t}\hat{y}+\cos\alpha_{t}\sin\alpha_{t}\sin\beta_{t}\hat{z})\tilde\Pi_t
\end{equation}
\begin{equation} 
\vec{H}^{p}_{t}= f^{p}_{t}\frac{ik_{1}^{2}\omega}{8\pi^{2}}
(\cos\alpha_{t}\sin^{2}\beta_{t}\hat{x}-\cos\alpha_{t}\cos\beta_{t}\sin\beta_{t}\hat{y})\tilde\Pi_t\,,
\end{equation}
and
\begin{equation} 
 \vec{H}^{s}_{t}= f^{s}_{t}\frac{ik_{1}^{2}\omega}{8\pi^{2}}
(\cos\alpha_{t}\cos^{2}\beta_{t}\hat{x}+\cos\alpha_{t}\cos\beta_{t}\sin\beta_{t}\hat{y}+\sin\alpha_{t}\cos\beta_{t}\hat{z})\tilde\Pi_t\,.
\end{equation}
where
\begin{equation}
\tilde\Pi_t = 
e^{ikz_{0}\cos{\alpha}}e^{ik_{1}(x\sin\alpha_{t}\cos\beta_{t}+y\sin\alpha_{t}\sin\beta_{t}-z\cos\alpha_{t})}
\end{equation} 
\begin{equation}
\vec k_t=k_{1}[\sin\alpha_{t}\cos\beta_{t}\hat x+ \sin\alpha_{t}\sin\beta_{t}\hat y
-\cos\alpha_{t} \hat z]\ ,
\end{equation}
 
We next impose 
the boundary conditions at the interface.
We shall assume $\mu=\mu_1$. 
We obtain
\begin{equation}
 k\sin\alpha=k_{1}\sin\alpha_{t}, \qquad \beta=\beta_{t}\,.
\end{equation}
\begin{equation}
f^{p}_{r} = \frac{k_{1}\cos\alpha-k\cos\alpha_{t}}{k_{1}\cos\alpha+k\cos\alpha_{t}} 
\end{equation}
\begin{equation}
f^{p}_{t}= \left(\frac{k}{k_{1}}\right)^{2}\left(\frac{1}{\cos\alpha_{t}}
\right)\frac{2k_{1}\cos^{2}\alpha}{k_{1}\cos\alpha+k\cos\alpha_{t}}\,.
\end{equation}
\begin{equation}
f^{s}_{r} = \frac{k\cos\alpha-k_{1}\cos\alpha_{t}}{k\cos\alpha+k_{1}\cos\alpha_{t}}
\end{equation}
and
\begin{equation}
f^{s}_{t}= \left(\frac{k}{k_{1}}\right)^{2}\frac{2k_{1}\cos\alpha}{k_{1}\cos\alpha_{t}+
k\cos\alpha}\,.
\end{equation}

Using the above Fresnel coefficients the $y$ components (H-Pol) of the reflected
and transmitted electric field at $y=0$ can be expressed as  
\begin{equation}
E_{r,y}={ik^{3}\over 8\epsilon\pi^{2}} \int_0^{2\pi}
\int_0^{{\pi\over 2}-i\infty} \tilde\Pi_{r}(f^{s}_{r}\cos^{2}\beta-f^{p}_{r}\cos^{2}\alpha\sin^{2}\beta) \sin\alpha d\alpha d\beta\,.
\label{eq:totalflat1}
\end{equation} 
\begin{equation}
E_{t,y}={ik_{1}^{3}\over 8\epsilon_{1}\pi^{2}} \int_0^{2\pi}
\int_0^{{\pi\over 2}-i\infty} \tilde\Pi_{t}(f^{s}_{t}\cos^{2}\beta_{t}+f^{p}_{t}\cos^{2}\alpha_{t}\sin^{2}\beta_{t}) \sin\alpha d\alpha d\beta \,.
\label{eq:totalflat2}
\end{equation} 
Using Eq. \ref{eq:totalflat1} we can compute the y-component of the total reflected field for a flat reflecting surface.
The resulting value of the reflection cofficient is found to be same as that for Fresnel reflection independent of frequency.

\section{Reflection and Transmission on a Spherical Surface using Local Plane Wave Approximation} 
\label{label:spherical}
\begin{figure}
\begin{centering}
\includegraphics[scale=0.8]{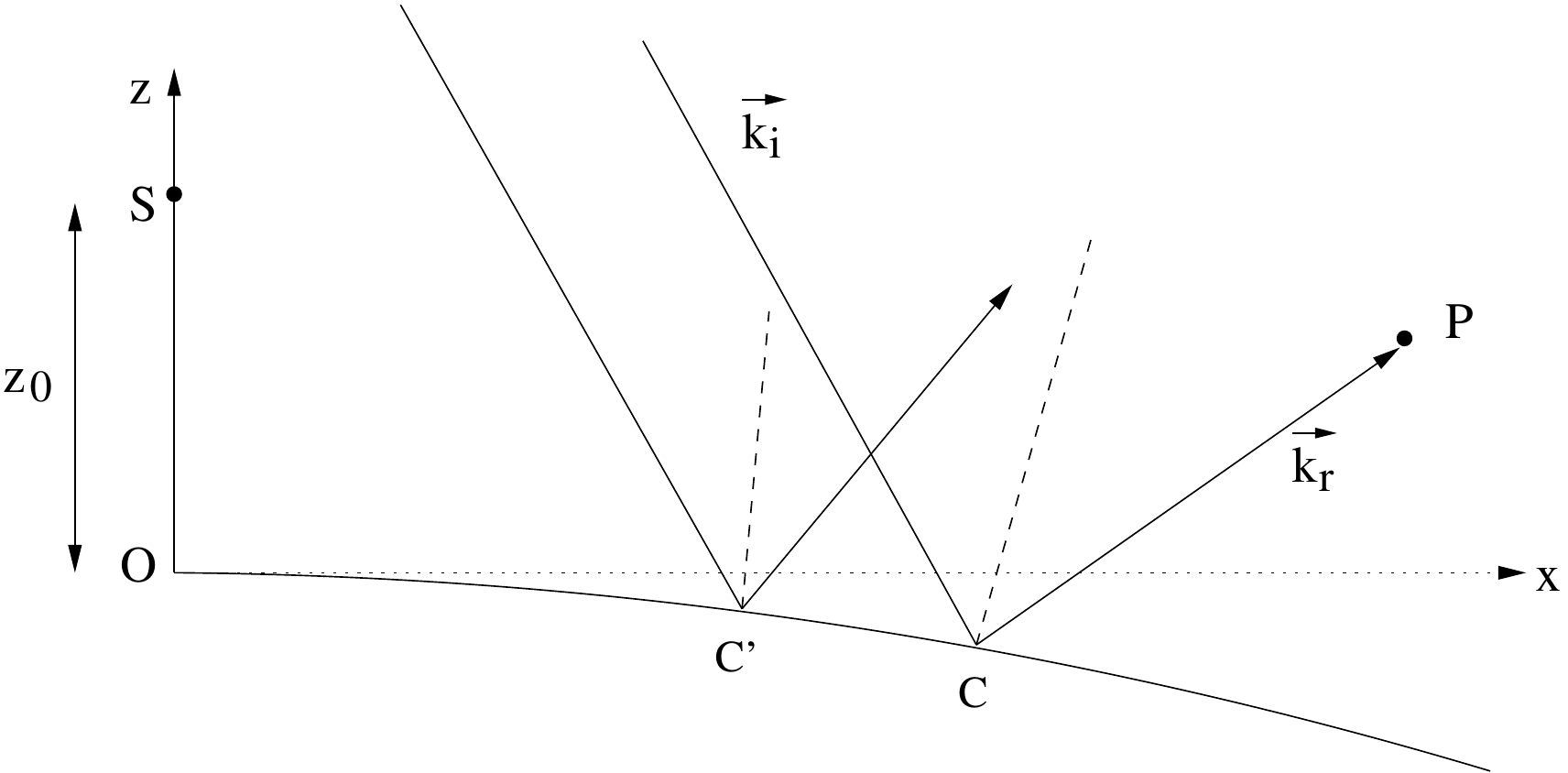}
\par\end{centering}
\caption{Here $S$ is the source and $P$ the detector. A plane wave with wave vector $\vec k_i$ reflects off the 
spherical surface. Here we show its reflection at two points $C$ and
$C'$. The reflected wave vector (such as $\vec k_r$) at 
any point (for example $C$) 
is obtained by constructing a tangent plane at that point. We obtain  
$\vec k_i$ by demanding that $\vec k_r$ points towards $P$. Notice that
different wave vectors corresponding to this incident plane wave which  
strike the surface at different points reflect off in different directions.
Hence the reflected wave is not a plane wave. We also show a wave vector
for this plane wave which reflects off at point $C'$.  
}
\label{fig:localgeometry}

\end{figure}

In this section we introduce a rigorous formalism to handle reflection of
spherical waves from a spherical surface. The radius of curvature is assumed
to be much larger than the wavelength. We again decompose the incident 
wave into plane waves (see Eq. \ref{eq:decomposition}). In an earlier 
calculation \cite{PhysRevD.98.042004}  we had assumed that the reflected wave 
corresponding to each incident plane wave is also a plane wave.  
This is a reasonable approximation since the curvature is very small.
However by direct comparison with HiCal observations the theoretical results
were found to disagree with data for small elevation angles.  
In the present paper we assume that the reflected wave is only locally
a plane wave. We explain this in Fig. \ref{fig:localgeometry}. Consider
an incident plane wave with wave vector $\vec k_i$ which reflects off
the curved surface. In the figure we have shown reflection from two points
$C$ and $C'$. For the point $C$ the reflected wave vector $\vec k_r$ can
be obtained by requiring that the reflection takes place from a
plane tangent to the surface at $C$. In the neighbourhood of the vector
$\vec k_r$ the wave is assumed to be locally a plane wave. The reflected
wave vector corresponding to point $C'$, however, points in a different 
direction. Hence globally the wave fronts are not plane. 
For a particular
incident plane wave we need to choose the incident wave vector 
for which the reflected wave vector $\vec k_r$ points towards the 
observation point $P$. 
We follow the same procedure for all plane waves and add the total 
contribution at the observation point.

\begin{figure}
\begin{centering}
\includegraphics[scale=0.52]{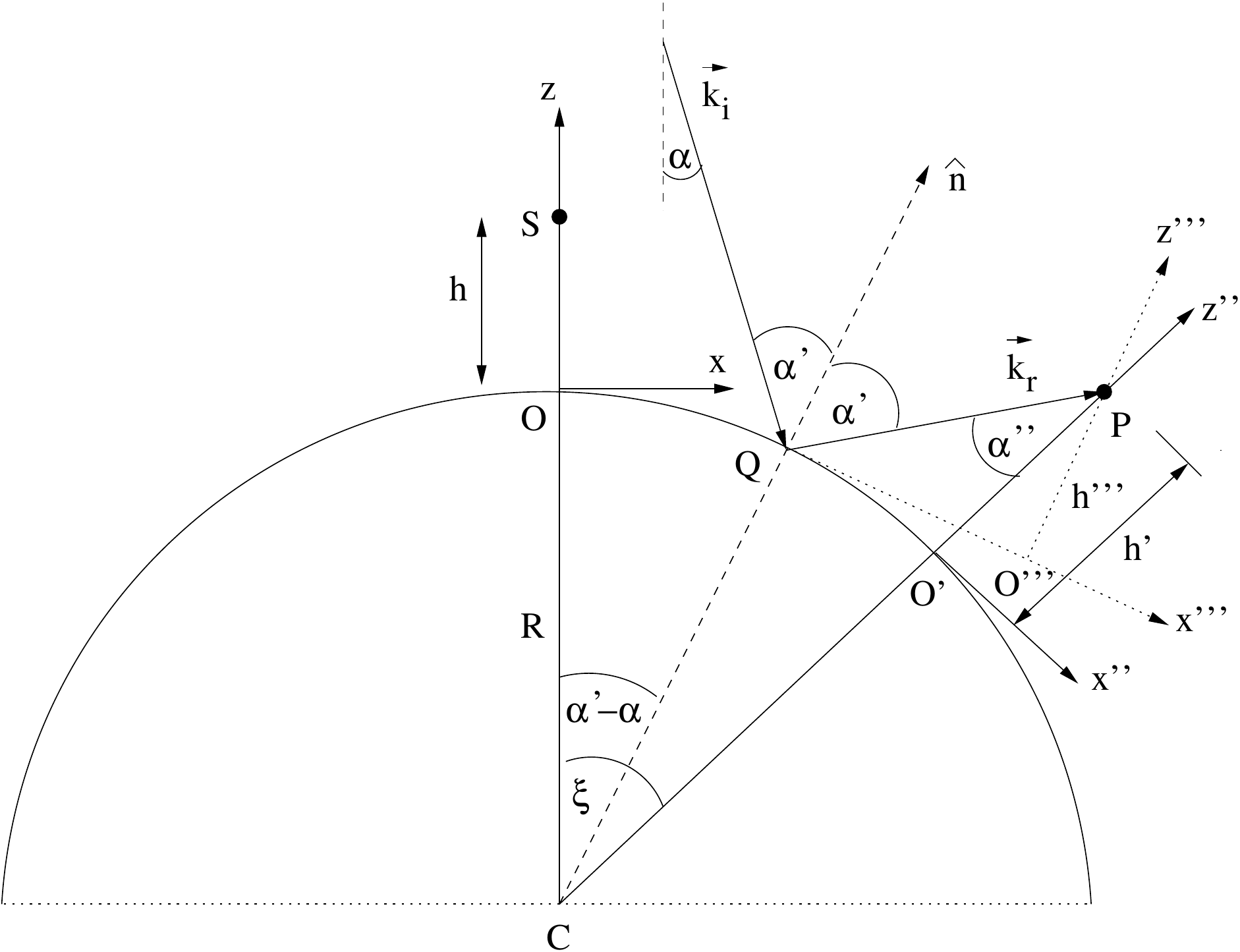}
\par\end{centering}
\caption{
The source is a horizontal dipole radiator located at $S(0,0,h)$ on the $z$ axis and the detector is at $P$. The point of reflection $Q$ for a particular plane wave is chosen such that
the reflected wave vector $\vec k_{r}$ points towards $P$. The incident wave vector is denoted by $\vec k_{i}$.    
Here $C$ is the centre of Earth and $R$ is the radius of Earth. 
 The position vector of $O^{\prime}$ with respect to $O$ is $\vec{r}_{0}$. 
  The normal to the surface at $Q$ is denoted by $\hat{n}$ which is parallel to the $z^{\prime\prime\prime}$ axes. 
 Angle of reflection and angle of transmission at point $Q$ are $\alpha^{\prime}$ and $\alpha^{\prime}_{t}$ (not shown in figure) respectively. 
}
\label{fig:local}
\end{figure}

We first need to determine the relationship between the angles $\alpha$
and $\alpha'$ for a particular plane wave, 
where $\alpha'$ is the reflection angle as shown in Fig. \ref{fig:local}. 
For this purpose
it is convenient to shift the coordinate system to $O'$.   
 The detector $P(x,0,z)$ is located vertically above
this point at altitude $h'$.  We identify a point $Q$ on the surface of Earth such that the reflected wave vector $\vec k_{r}$ corresponding to the incident 
wave vector $\vec k_{i}$ at this point is directed towards the detector at $P$. We start with a reflected wave vector $\vec{k}_{r}$ and find the corresponding incident wave vector $\vec{k}_{i}$ as shown in Fig. \ref{fig:local}.

For a given point of observation $P(x,0,z)$ the angle $\xi$, defined in 
Fig. \ref{fig:local}, is given by  $\tan\xi= \frac{x}{R+z}$, where R is the radius of Earth. We first rotate our coordinate system about the $y$ axis by angle $\xi$ and obtain coordinate system $x^{\prime\prime}-y^{\prime\prime}-z^{\prime\prime}$. The $z^{\prime\prime}$ axis meets the surface of Earth at the point $O^{\prime}$. We choose $O^{\prime}$ as our reference point for further calculations. The rotation matrix corresponding to this is  
\begin{equation}
 R_{y}(\xi)=\left(\begin{array}{ccc}\cos\xi & 0&-\sin\xi  \\ 0 & 1 & 0\\\sin\xi & 0&\cos\xi  \end{array}\right)
\end{equation}
The unit vectors in this new coordinate system are related to the unit vectors in $x-y-z$ by 
\begin{align}
\hat{x}^{\prime\prime}= \cos\xi \hat{x}-\sin\xi \hat{z}\nonumber \\
\hat{y}^{\prime\prime}= \hat{y} \nonumber  \\
\hat{z}^{\prime\prime}= \sin\xi \hat{x}+\cos\xi \hat{z}
\end{align}
Next we rotate the coordinate system ($x^{\prime\prime}-y^{\prime\prime}-z^{\prime\prime}$)  about the $z^{\prime\prime}$ axis by an angle $\tilde{\beta}$. This leads to the rotation matrix as, 
\begin{equation}
 R_{z^{\prime\prime}}(\tilde{\beta})=\left(\begin{array}{ccc}\cos\tilde{\beta} & \sin\tilde{\beta}& 0  \\ -\sin\tilde{\beta} & \cos\tilde{\beta} & 0\\0 & 0&1  \end{array}\right)
\end{equation}
Note that the rotation about $z^{\prime\prime}$ axis by angle $\tilde{\beta}$ 
 enables us to get the contribution from all locally plane waves that 
reach the detector at $P$ by use of cylindrical symmetry.  
Due to the second rotation, we finally obtain our coordinate system $(x'-y'-z')$ such that \\
\begin{align}
\hat{x}^{\prime}= \cos\tilde{\beta} \hat{x}^{\prime\prime}+\sin\tilde{\beta} \hat{y}^{\prime\prime}  \nonumber \\
\hat{y}^{\prime}= -\sin\tilde{\beta}\hat{x}^{\prime\prime} +\cos\tilde{\beta}\hat{y}^{\prime\prime} \nonumber  \\
\hat{z}^{\prime}= \hat{z}^{\prime\prime}
\end{align}
Now the overall rotation matrix is given by, \
 $Rot_{1}=  R_{z^{\prime\prime}}(\tilde{\beta}) R_{y}(\xi)$\
\begin{equation} \label{two rotation}
Rot_{1} = \left(\begin{array}{ccc}\cos\xi\cos\tilde{\beta} & \sin\tilde{\beta} & -\sin\xi\cos\tilde{\beta}\\ -\cos\xi\sin\tilde{\beta} & \cos\tilde{\beta} & \sin\xi\sin\tilde{\beta} \\ 
\sin\xi &0 & \cos\xi \end{array}\right)
\end{equation}

Next we write the incident wave vector in the new coordinate system as, 
\begin{equation}
 \vec{k}^{\prime}_{i}= Rot_{1} \cdot \vec{k}_{i} = k_{x}^{\prime}\hat{x}^{\prime}+ k_{y}^{\prime}\hat{y}^{\prime}+ k_{z}^{\prime}\hat{z}^{\prime} 
\end{equation}
Using \ref{eq:kinc} and \ref{two rotation}  we obtain the components of incident wave vector in $x^{\prime}-y^{\prime}-z^{\prime}$ coordinate system as
\begin{equation}
k_{x}^{\prime}=k((\sin\alpha\cos\beta\cos\xi+\cos\alpha\sin\xi)\cos\tilde{\beta}+\sin\alpha\sin\beta\sin\tilde{\beta})\nonumber
\end{equation} 
\begin{equation}
k_{y}^{\prime}=k((-\sin\alpha\cos\beta\cos\xi-\cos\alpha\sin\xi)\sin\tilde{\beta}+\sin\alpha\sin\beta\cos\tilde{\beta})\nonumber
\end{equation} 
\begin{equation}
k_{z}^{\prime}=k((-\sin\alpha\cos\beta\sin\xi+\cos\alpha\cos\xi)\sin\tilde{\beta}+\sin\alpha\sin\beta\cos\tilde{\beta})
\label{eq:k1}
\end{equation}
Since $\vec{k}^{\prime}_{i}$ lies in the $x^{\prime}-z^{\prime}$ plane which is our plane of incidence, we must set $y^{\prime}$ component of  $\vec{k}^{\prime}_{i}$ as zero. This leads to the expression of $\tilde{\beta}$ as
\begin{equation}
\tan\tilde{\beta}= \frac{\sin\alpha\sin\beta}{\sin\alpha\cos\beta\cos\xi+\cos\alpha\sin\xi}
\end{equation}
We introduce a new parameter $\tilde{\alpha}$ such that 
\begin{equation}
\cos\tilde{\alpha}=(\cos\alpha\cos\xi-\sin\alpha\cos\beta\sin\xi)
\label{eq:tilde_alpha1}
\end{equation}
and 
\begin{equation}
\sin\tilde{\alpha}=((\sin\alpha\cos\xi\cos\beta+\cos\alpha\sin\xi)^{2}+(\sin\alpha\sin\beta)^{2})^{\frac{1}{2}}
\label{eq:tilde_alpha2}
\end{equation}
Using \ref{eq:k1}$-$\ref{eq:tilde_alpha2} we simplify the expression for incident wave vector $\vec{k}^{\prime}_{i}$ and express it as
\begin{equation}
 \vec{k}^{\prime}_{i}= k(\sin\tilde{\alpha}\hat{x}^{\prime}-\cos\tilde{\alpha}\hat{z}^{\prime})
\label{eqn:kinc_prime}
\end{equation}
The reflected wave vector in the new coordinate system is given by,
\begin{equation}
 \vec{k}^{\prime}_{r}= Rot_{1} \cdot \vec{k}_{r} = k(\sin\alpha^{\prime\prime}\hat{x}^{\prime}+\cos\alpha^{\prime\prime}\hat{z}^{\prime})
 \label{eqn:kref_prime}
 \end{equation}
Similarly, the transmitted wave vector in the new coordinate system is given by,
\begin{equation}
 \vec{k}^{\prime}_{t}= Rot_{1} \cdot \vec{k}_{t} = k_1(\sin\tilde{\alpha}_{t}\hat{x}^{\prime}-\cos\tilde{\alpha}_{t}\hat{z}^{\prime})
\label{eqn:ktrans_prime}
\end{equation}
The normal to the spherical surface at $Q$ is given by
\begin{equation}
\hat{n}=\cos(\xi-(\alpha^{\prime}-\alpha))\hat{z}^{\prime}-\sin(\xi-(\alpha^{\prime}-\alpha))\hat{x}^{\prime} 
\end{equation}
From \ref{eqn:kinc_prime}$-$\ref{eqn:ktrans_prime}, we see that the incident, reflected and transmitted wave vectors lie in the same plane. The normal $\hat{n}$ must also lie in the $x^{\prime}-z^{\prime}$ plane to satisfy the laws of reflection at the interface between air and ice. 
 Using the condition $-\hat{n}\cdot\vec{k}^{\prime}_{i}= \hat{n}\cdot \vec{k}^{\prime}_{r}$, we get a relation between angle of reflection $\alpha^{\prime}$ and angle $\alpha$ as
 \begin{equation}
 \alpha^{\prime\prime}= -2\xi+2\alpha^{\prime}-2\alpha+\tilde{\alpha}
 \label{eq:alpha1}
 \end{equation}
 From the geometry (see Fig. \ref{fig:local} ) we derive
 \begin{equation}
 \sin\alpha^{\prime}= \frac{R+h^{\prime}}{R} \sin\alpha^{\prime\prime}
 \label{eq:alpha2}
 \end{equation}
 Using \ref{eq:tilde_alpha1}, \ref{eq:alpha1} and \ref{eq:alpha2} we numerically solve this equation in order to determine $\alpha^{\prime}$ as a function of $\alpha$ for a given observation point $P$ (see Fig.\ref{fig:local}). The reference point $O^{\prime}$ is 
located at $\vec{r}_{0}$ from the origin $O(0,0,0)$. The point of reflection $Q$ and the observation point $P$ are located at $\vec{r}_s^{\prime}$ and $\vec{r}^{\prime}$ respectively with respect to $O^{\prime}$. From the geometry (see Fig. \ref{fig:local}) we obtain the expression for $\vec{r}_{0}$, $\vec{r}^{\prime}_{s}$ and $\vec{r}^{\prime}$ as
 
\begin{equation}
\vec{r}^{\prime}=h^{\prime}\hat{z}^{\prime\prime}=h^{\prime}\hat{z}^{\prime}
\end{equation}
\begin{equation}
R\hat{z}+\vec{r}_{0}=\vec{r}_{2}= R(\sin\xi\hat{x}+\cos\xi\hat{z}) \nonumber
\end{equation} 
or
\begin{equation}
\vec{r}_{0}= R(\cos\xi-1) \hat{z}+R\sin\xi \hat{x}
\end{equation}
and 
\begin{equation}
\begin{aligned}
\vec{r}^{\prime}_{s}=\vec{r}_1-\vec{r}_2 = -R\sin(\xi-\alpha^{\prime}+\alpha)\hat{x}^{\prime\prime}-R(1-\cos(\xi-\alpha^{\prime}+\alpha))\hat{z}^{\prime\prime}
\end{aligned}
\label{eq:rsprime1}
\end{equation}
For $\tilde{\beta}$=$\beta$=$0$, we get 
\begin{equation}
\begin{aligned}
\xi=2\alpha^{\prime}-\alpha-\alpha^{\prime\prime} \nonumber
\end{aligned}
\end{equation}
or
\begin{equation}
\begin{aligned}
\xi-\alpha^{\prime}+\alpha=\alpha^{\prime}-\alpha^{\prime\prime} .
\end{aligned}
\label{eq:rsprime2}
\end{equation}
Using Eq. \ref{eq:rsprime1} and \ref{eq:rsprime2}, we obtain the expression for $\vec{r}^{\prime}_{s}$ as
\begin{equation}
\begin{aligned}
\vec{r}^{\prime}_{s}=-R\sin(\alpha^{\prime}-\alpha^{\prime\prime})\hat{x}^{\prime}-R(1-\cos(\alpha^{\prime}-\alpha^{\prime\prime}))\hat{z}^{\prime} ,
\end{aligned}
\end{equation}
which is valid for any $\tilde \beta$. 
The exponential factor for the incident plane wave is derived for the spherical geometry using the same method as in the case of flat geometry.
We express it as,
\begin{eqnarray}
\tilde\Pi_{S,i}  &=& \exp\left[{i\vec{k}^{\prime}_{i}\cdot (\vec r^{\prime} +\vec{r}_{0}-h\hat{z}) }\right]\nonumber \\
&=& \exp\left[{ik(x\sin\alpha\cos\beta+y\sin\alpha\sin\beta+(h-z)\cos\alpha}\right] .
\end{eqnarray}
From geometry (Fig. \ref{fig:local}) we find the expression for $h^{\prime}$ as  
\begin{equation}
h^{\prime}=R\left[\left(1+\frac{z}{R}\right)^{2}+\left(\frac{x}{R}\right)^{2}\right]^{\frac{1}{2}}-R .
\end{equation}
Now we obtain the exponential factor for the reflected wave by applying the 
boundary conditions at the point $Q$ whose position vector with respect to $O'$ is $\vec{r}_{s}^{\prime}$.
\begin{equation}
\vec{k}^{\prime}_{i}\cdot \vec{r}_{s}^{\,\prime}+\vec{k}^{\prime}_{i}\cdot(
\vec{r}_{0}-h\hat{z}) = \vec{k}^{\prime}_{r}\cdot \vec{r}_{s}^{\,\prime}+D \\
\end{equation}
This fixes the value of $D$ and
the resulting expression for $\tilde\Pi_{S,r}$ is given by
\begin{eqnarray}
\tilde\Pi_{S,r}  &=& \exp[i(\vec{k}^{\prime}_{r}\cdot \vec{r}^{\prime}+D)]
\nonumber\\
&=&\exp[i(\vec{k}^{\prime}_{r}\cdot \vec{r}^{\prime}+\vec{k}^{\prime}_{i}\cdot \vec{r}_s^{\prime}+\vec{k}^{\prime}_{i}\cdot(\vec{r_{0}}-h\hat{z})-\vec{k}^{\prime}_{r}\cdot \vec{r}_s^{\prime})]  . \nonumber 
\label{eq:Pisr}
\end{eqnarray}
This leads to
\begin{equation}
\begin{aligned}
\tilde\Pi_{S,r}  =
\exp[ik((R+h^{\prime})\cos\alpha^{\prime\prime}+(R+h)\cos\alpha-R\cos\alpha^{\prime}-R\cos(\alpha^{\prime}-\alpha^{\prime\prime}-\tilde{\alpha}))] 
\end{aligned}
\end{equation}
Similarly, we determine the exponential factor for the transmitted wave from the geometry (see Fig. \ref{fig:local} ) . We express it as
\begin{equation}
\tilde\Pi_{S,t} = e^{i\vec{k}^{\prime}_{i}\cdot\vec{\Delta}^{\prime}}e^{i\vec{k}^{\prime}_{t}\cdot \vec r'}
\end{equation} 
where $e^{i\vec{k}^{\prime}_{i}\cdot\vec{\Delta}^{\prime}}$ is the constant term appearing in both $\tilde\Pi_{S,i}$ and $\tilde\Pi_{S,t}$. As in the case of flat geometry, this term is proportional to k and not $k_{1}$. 
We write the electric field components in the ($x^{\prime}-y^{\prime}-z^{\prime}$) coordinate system as
\begin{equation}
 \vec{E}^{\prime}_{i}= Rot_{1} \cdot \vec{E}_{i}= \frac{ik^{3}}{8\epsilon\pi^{2}} \tilde\Pi_{S,i} [\sin\tilde\beta\cos^{2}\tilde\alpha\hat{x}^{\prime}+\cos\tilde\beta\hat{y}^{\prime}+\cos\tilde\alpha\sin\alpha\sin\beta\hat{z}^{\prime}] \nonumber
\end{equation}
\begin{equation}
 \vec{H}^{\prime}_{i}= Rot_{1} \cdot \vec{H}_{i}= \frac{ik^{2}\omega}{8\pi^{2}}\tilde\Pi_{S,i} [\cos\tilde\alpha\cos\tilde\beta\hat{x}^{\prime}-\cos\tilde\alpha\sin\tilde\beta\hat{y}^{\prime}+\sin\tilde\alpha\cos\tilde\beta\hat{z}^{\prime}]
\end{equation}
\subsection{Deriving Reflection and Transmission Coefficients and 
Comparison with HiCal Data }

As explained in earlier sections the basic idea of the local plane wave
approximation is that we demand that the reflected wave vector 
$\vec k_r$ corresponding to any incident plane wave points towards the detector located at $P$. 
For all such vectors we determine the incident wave vectors $\vec k_i$
and eventually need to integrate over the contributions from all the plane
waves. In order to determine the reflected and transmitted waves we 
follow the same procedure as before \cite{PhysRevD.98.042004}. For each plane
wave we determine the tangent plane at the point of reflection $Q$. 
The tangent
plane acts as the flat reflecting surface and now we can use the
procedure for flat surface to determine the Fresnel coefficients. 
As in \cite{PhysRevD.98.042004} we transform to a coordinate system $x'''-y'''-z'''$
in which this surface is the $x'''-y'''$ plane. 
 Using the reflection coefficients, we compute the electric and magnetic field components for each plane wave in the new coordinate system. 
Since the new coordinate system is not fixed rather it depends on the plane
wave under consideration, we transform our fields back to the original frame and integrate over all plane waves to get the total field. 

For a given plane wave, let the point $Q$ be located at $(x_s,y_{s},z_{s})$ with respect to origin $O(0,0,0)$.
 We identify the tangent plane at this point and choose a new coordinate system ($x'''-y'''-z'''$) such that it satisfies the following conditions: 
\begin{itemize} 
\item[1.] The coordinates of $Q$ in this system are $(x^{\prime\prime\prime}_{s},0,0)$. 

\item[2.]  The observation point $P$ lies on the $z^{\prime\prime\prime}$ 
axis at a height $h^{\prime\prime\prime}$ above the new reference point 
$O^{\prime\prime\prime}$.

\item[3.]  The unit vector $\hat n$, normal  to the tangent plane at $Q$ is parallel to the $z^{\prime\prime\prime}$ axis  (see Fig. \ref{fig:local} ). 
\end{itemize}

 We have 
\begin{equation}
 \hat{n}=-\sin\psi\hat{x}^{\prime}+\cos\psi\hat{z}^{\prime} \\ 
\end{equation}
where
\begin{equation}
\psi= \xi -\alpha^{\prime}+\alpha 
\end{equation}
The final rotation matrix in order to transform from $x'-y'-z'$ to 
 $x^{\prime\prime\prime}-y^{\prime\prime\prime}-z^{\prime\prime\prime}$ coordinate system is given by, 
\begin{equation} \label{3rd rotation}
R_{y^{\prime}}(\psi) = \left(\begin{array}{ccc}\cos\psi & 0 &\sin\psi\\ 0 & 1 & 0 \\ 
-\sin\psi & 0 &\cos\psi\\  \end{array}\right)
\end{equation}
The unit vectors in this coordinate system are 
\begin{eqnarray}
\hat{x}^{\prime\prime\prime} &=& \cos\psi\hat{x}^{\prime}+\sin\psi\hat{z}^{\prime}   \nonumber\\ 
\hat{y}^{\prime\prime\prime} &=& \hat{y}^{\prime}  \nonumber\\     
\hat{z}^{\prime\prime\prime} &=& -\sin\psi\hat{x}^{\prime}+\cos\psi\hat{z}^{\prime} 
\end{eqnarray}
Incorporating all three rotations, we obtain the full rotation matrix using Eq.(\ref{two rotation}) and Eq.(\ref{3rd rotation}) as
\begin{equation}
Rot=  R_{y^{\prime}}(\psi)Rot_{1}=  R_{y^{\prime}}(\psi) R_{z^{\prime\prime}}(\tilde{\beta}) R_{y}(\xi)
\end{equation}
We find the incident, reflected and transmitted wave vectors in the new coordinate system as
\begin{equation}
\vec{k}^{\prime\prime\prime}_{i} = k[\sin(\tilde\alpha-\psi)\hat{x}^{\prime\prime\prime}-\cos(\tilde\alpha-\psi)\hat{z}^{\prime\prime\prime}] \nonumber
\end{equation}
\begin{equation}
\vec{k}^{\prime\prime\prime}_{r} = k[\sin(\alpha^{\prime\prime}+\psi)\hat{x}^{\prime\prime\prime}+\cos(\alpha^{\prime\prime}+\psi)\hat{z}^{\prime\prime\prime}] \nonumber
\end{equation}
\begin{equation}
\vec{k}^{\prime\prime\prime}_{t} = k_1[\sin(\tilde\alpha_{t}-\psi)\hat{x}^{\prime\prime\prime}-\cos(\tilde\alpha_{t}-\psi)\hat{z}^{\prime\prime\prime}]
\label{eq:final_k} 
\end{equation}
We have simplified the expression for transmitted wave vector $\vec{k}^{\prime\prime\prime}_{t}$ using 
\begin{equation}
\cos\tilde{\alpha_{t}}=(\cos\alpha_{t}\cos\xi-\sin\alpha_{t}\cos\beta_{t}\sin\xi) 
\end{equation}

The electric and magnetic field expressions in the ($x^{\prime\prime\prime}-y^{\prime\prime\prime}-z^{\prime\prime\prime}$) are obtained as
\begin{eqnarray}
\vec{E}^{\prime\prime\prime}_{i} &=&  
R_{y^{\prime}}(\psi)\vec{E}^{\prime}_{i}\nonumber\\ 
&=&\frac{ik^{3}}{8\epsilon\pi^{2}}\tilde\Pi_{S,i}  [(\sin\tilde\beta\cos^{2}\tilde\alpha\cos\psi+\sin\alpha\cos\tilde\alpha\sin\beta\sin\psi)\hat{x}^{\prime\prime\prime}+\cos\tilde\beta\hat{y}^{\prime\prime\prime}\nonumber\\
&+&(\sin\alpha\cos\tilde\alpha\sin\beta\cos\psi-\cos^{2}\tilde\alpha\sin\tilde\beta\sin\psi)\hat{z}^{\prime\prime\prime}]
\end{eqnarray}

\begin{eqnarray}
\vec{H}^{\prime\prime\prime}_{i} &=& 
 R_{y^{\prime}}(\psi)\vec{H}^{\prime}_{i}\nonumber\\
&=&\frac{ik^{2}\omega}{8\pi^{2}}\tilde\Pi_{S,i} [\cos\tilde\beta\cos(\tilde\alpha-\psi)\hat{x}^{\prime\prime\prime}-\cos\tilde\alpha\sin\tilde\beta\hat{y}^{\prime\prime\prime}\nonumber\\
&+&\cos\tilde\beta\sin(\tilde\alpha-\psi)\hat{z}^{\prime\prime\prime}]
\end{eqnarray}
Now we use the same method as in the case of flat geometry to find the $s$ and $p$ components of $E^{\prime\prime\prime}_{q}$ and  $H^{\prime\prime\prime}_{q}$  (where the subscript $q$ designates the incident, reflected or transmitted waves).

We derive the reflection and transmission coefficients by imposing boundary conditions at $Q$ i.e. $z^{\prime\prime\prime}_{s}=0$. We find the unit vector normal to the plane of incidence corresponding to wave vector $\vec{k}^{\prime\prime\prime}_{i}$ as \\
\begin{equation}
 \hat{\eta}= l\hat{x}^{\prime\prime\prime}+m\hat{y}^{\prime\prime\prime}+n\hat{z}^{\prime\prime\prime}
 \end{equation}
 The vector $\vec{k}^{\prime\prime\prime}_{i}$ and $\hat{z}^{\prime\prime\prime}$ lie in the plane of incidence and hence are perpendicular to $\hat{\eta}$. This implies that $n=0$ and $(l\hat{x}^{\prime\prime\prime}+m\hat{y}^{\prime\prime\prime}+n\hat{z}^{\prime\prime\prime})\cdot \vec{k}^{\prime\prime\prime}_{i}=0$. The resulting unit vector $\hat{\eta}$ perpendicular to the plane of incidence is given by, 
\begin{equation}
\hat{\eta}= \hat{y}^{\prime\prime\prime}
\end{equation}
Now we write the $s$ and $p$ components of incident electric field as
\begin{equation}
\vec{E}^{\prime\prime\prime(s)}_{i}= (\vec{E}^{\prime\prime\prime}_{i}\cdot \hat{\eta})\hat{\eta}=   
\frac{ik^{3}}{8\epsilon\pi^{2}}\tilde\Pi_{S,i}  \cos\tilde\beta\hat{y}^{\prime\prime\prime} \nonumber
\end{equation}
\begin{eqnarray}
\vec{E}^{\prime\prime\prime(p)}_{i}= \vec{E}^{\prime\prime\prime}_{i}-\vec{E}^{\prime\prime\prime(s)}_{i}&=&\frac{ik^{3}}{8\epsilon\pi^{2}}\tilde\Pi_{S,i}[(\sin\tilde\beta\cos^{2}\tilde\alpha\cos\psi+\sin\alpha\cos\tilde\alpha\sin\beta\sin\psi)\hat{x}^{\prime\prime\prime}\nonumber\\
&+&(\sin\alpha\cos\tilde\alpha\sin\beta\cos\psi-\cos^{2}\tilde\alpha\sin\tilde\beta\sin\psi)\hat{z}^{\prime\prime\prime}]
\end{eqnarray}
Similarly the incident magnetic field components can be written as
\begin{equation}
\vec{H}^{\prime\prime\prime(p)}_{i}= (\vec{H}^{\prime\prime\prime}_{i}\cdot \hat{\eta})\hat{\eta}=\frac{ik^{2}\omega}{8\pi^{2}}\tilde\Pi_{S,i} 
  [-\cos\tilde\alpha\sin\tilde\beta\hat{y}^{\prime\prime\prime}]\nonumber
\end{equation}

\begin{equation}
\vec{H}^{\prime\prime\prime(s)}_{i}= \vec{H}^{\prime\prime\prime}_{i}-\vec{H}^{\prime\prime\prime(p)}_{i}
=\frac{ik^{2}\omega}{8\pi^{2}}\tilde\Pi_{S,i} [\cos\tilde\beta\cos(\tilde\alpha-\psi)\hat{x}^{\prime\prime\prime}+\cos\tilde\beta\sin(\tilde\alpha-\psi)\hat{z}^{\prime\prime\prime}]
\end{equation}
The s and p components of reflected electric field are obtained as
\begin{equation}
\vec{E}^{\prime\prime\prime(s)}_{r}= f^{\prime (s)}_{r} 
\frac{ik^{3}}{8\epsilon\pi^{2}}\tilde\Pi_{S,r} [ \cos\tilde\beta\hat{y}^{\prime\prime\prime}],  \nonumber
\end{equation}
\begin{eqnarray}
 \vec{E}^{\prime\prime\prime(p)}_{r} &=& f^{\prime (p)}_{r} \frac{ik^3}{8\epsilon\pi^{2}}\,
\tilde \Pi_{S,r}\,
[-(\sin\tilde\beta\cos^{2}\tilde\alpha\cos\psi+\sin\alpha\cos\tilde\alpha\sin\beta\sin\psi)\hat{x}^{\prime\prime\prime}\nonumber\\
&+&(\sin\alpha\cos\tilde\alpha\sin\beta\cos\psi-\cos^{2}\tilde\alpha\sin\tilde\beta\sin\psi)\hat{z}^{\prime\prime\prime}]
\end{eqnarray}
Similarly, for the reflected magnetic field components we write
\begin{equation}
\vec{H}^{\prime\prime\prime(p)}_{r}= f^{\prime (p)}_{r}  \frac{ik^{2}\omega}{8\pi^{2}}\tilde\Pi_{S,r} [-\cos\tilde\alpha\sin\tilde\beta\hat{y}^{\prime\prime\prime}]\nonumber
\end{equation}
\begin{equation}
\vec{H}^{\prime\prime\prime(s)}_{r}= f^{\prime (s)}_{r} \frac{ik^{2}\omega}{8\pi^{2}}\tilde\Pi_{S,r}  [-\cos\tilde\beta\cos(\tilde\alpha-\psi)\hat{x}^{\prime\prime\prime}+\cos\tilde\beta\sin(\tilde\alpha-\psi)\hat{z}^{\prime\prime\prime}]
\end{equation}
where, $f^{\prime (s)}_{r}$ and $f^{\prime (p)}_{r}$ are fresnel coefficients corresponding to s and p component of reflected fields.

The corresponding transmitted field $\vec{E}^{\prime\prime\prime(s)}_{t}$,$\vec{E}^{\prime\prime\prime(p)}_{t}$,$\vec{H}^{\prime\prime\prime(s)}_{t}$ and $\vec{H}^{\prime\prime\prime(p)}_{t}$ can be written as
\begin{equation}
\vec{E}^{\prime\prime\prime(s)}_{t}=    f^{\prime (s)}_{t} \frac{ik_{1}^3}{8\epsilon_{1}\pi^{2}}\,
\tilde \Pi_{S,t}\, [\cos\tilde\beta_{t} \hat{y}^{\prime\prime\prime}]\nonumber
\end{equation}
\begin{eqnarray}
\vec{E}^{\prime\prime\prime(p)}_{t} &=& \vec{E}^{\prime\prime\prime}_{t}-\vec{E}^{\prime\prime\prime(s)}_{t}\nonumber\\
&=&f^{\prime (p)}_{t}\frac{ik_{1}^3}{8\epsilon_{1}\pi^{2}}\,
\tilde \Pi_{S,t}\,
[(\sin\tilde\beta_{t}\cos^{2}\tilde\alpha_{t}\cos\psi+\sin\alpha_{t}\cos\tilde\alpha_{t}\sin\beta_t \sin\psi)\hat{x}^{\prime\prime\prime}
\nonumber\\
&+&(\sin\alpha_{t}\cos\tilde\alpha_{t}\sin\beta_{t}\cos\psi-\cos^{2}\tilde\alpha_{t}\sin\tilde\beta_{t}\sin\psi)\hat{z}^{\prime\prime\prime}]
\end{eqnarray}
Similarly the transmitted magnetic field components can be written as 
\begin{equation}
\vec{H}^{\prime\prime\prime(p)}_{t}= f^{\prime (p)}_{t}\frac{ik_{1}^2\omega}{8\pi^{2}}\,
\tilde \Pi_{S,t}\, [-\cos\tilde\alpha_{t}\sin\tilde\beta_{t}\hat{y}^{\prime\prime\prime}]\nonumber
\end{equation}
\begin{eqnarray}
\vec{H}^{\prime\prime\prime(s)}_{t}= \vec{H}^{\prime\prime\prime}_{t}-\vec{H}^{\prime\prime\prime(p)}_{t}
&=&f^{\prime (s)}_{t} \frac{ik_{1}^2\omega}{8\pi^{2}}\,
\tilde \Pi_{S,t}\, [\cos\tilde\beta_{t}\cos(\tilde\alpha_{t}-\psi)\hat{x}^{\prime\prime\prime}\nonumber\\
&+&\cos\tilde\beta_{t}\sin(\tilde\alpha_{t}-\psi)\hat{z}^{\prime\prime\prime}]
\end{eqnarray}
We impose the boundary conditions at $z^{\prime\prime\prime}_{s}=0$ on each components in order to determine the reflection coefficients. We use the same procedure as described in \cite{PhysRevD.98.042004}. The exponential factors lead to the standard conditions:
\begin{eqnarray}
 k\sin(\tilde\alpha-\psi)&=&k_{1}\sin(\tilde\alpha_{t}-\psi) \nonumber\\
\tilde\beta_{t}&=& \tilde\beta
\end{eqnarray}
The continuity of electric field components $\parallel$ to the surface imply that 
$$\vec{E}^{\prime\prime\prime (p)}_{t,x}=\vec{E}^{\prime\prime\prime (p)}_{i,x}+\vec{E}^{\prime\prime\prime (p)}_{r,x}$$
The components $\perp$ to the surface follow:
 $$\epsilon_{1}\vec{E}^{\prime\prime\prime (p)}_{t,z}=\epsilon[\vec{E}^{\prime\prime\prime (p)}_{i,z}+
\vec{E}^{\prime\prime\prime (p)}_{r,z}]$$
The component of magnetic field $\perp$ to the surface are continuous at
the interface and the parallel components satisfy,
 $$\mu_{1}\vec{H}^{\prime\prime\prime (p)}_{t,y}=\mu\left[\vec{H}^{\prime\prime\prime (p)}_{i,y}+
\vec{H}^{\prime\prime\prime (p)}_{r,y}\right]\,.$$ 
Here we shall assume $\mu_{1}=\mu$.
These conditions lead to:
\begin{equation}
  (1-f^{\prime (p) }_{r})= f^{\prime (p) }_{t}\frac{k_{1}}{k}\frac{\cos\tilde\alpha_{t}\cos(\tilde\alpha_{t}-\psi)}{\cos\tilde\alpha\cos(\tilde\alpha-\psi)}
\label{eq:Fresnelp11}
\end{equation}
\begin{equation}
 (1+f^{\prime (p) }_{r})= f^{\prime (p) }_{t}\frac{k_{1}^{3}}{k^{3}}\frac{\cos\tilde\alpha_{t}\sin(\tilde\alpha_{t}-\psi)}{\cos\tilde\alpha\sin(\tilde\alpha-\psi)}
 =  f^{\prime (p) }_{t}\frac{k_{1}^{2}}{k^{2}}\frac{\cos\tilde\alpha_{t}}{\cos\tilde\alpha}
\label{eq:Fresnelp22}  
\end{equation}
 Solving Eqs. \ref{eq:Fresnelp11} and \ref{eq:Fresnelp22} we obtain 
\begin{equation}
  f^{\prime (p) }_{r}= \frac{k_{1}\cos(\tilde\alpha-\psi)-k\cos(\tilde\alpha_{t}-\psi)}{k_{1}\cos(\tilde\alpha-\psi)+k\cos(\tilde\alpha_{t}-\psi)} \nonumber ,
  \end{equation}
and 
\begin{equation}
f^{\prime (p)}_{t}= \left(\frac{k}{k_{1}}\right)^{2}\left(\frac{1}{\cos\tilde\alpha_{t}}
\right)\frac{2k_{1}\cos(\tilde\alpha-\psi)\cos\tilde\alpha}{k_{1}\cos(\tilde\alpha-\psi)+k\cos(\tilde\alpha_{t}-\psi)}
\end{equation}
Next we impose boundary conditions on the components $\perp$ to the plane of incidence. These leads to
$$\vec{E}^{\prime\prime\prime (s)}_{t,y}=\vec{E}^{\prime\prime\prime (s)}_{i,y}+\vec{E}^{\prime\prime\prime (s)}_{r,y}$$
 and $$\mu_{1}\vec{H}^{\prime\prime\prime (p)}_{t,x}=\mu\left[\vec{H}^{\prime\prime\prime (p)}_{i,x }+
\vec{H}^{\prime\prime\prime (p)}_{r,x}\right]\,$$
The $z$ component of the electric field  does not lead to a new condition.
These conditions imply
\begin{equation}
 (1+f^{\prime (s)}_{r})= f^{\prime (s)}_{t}\frac{k_{1}}{k}
 \label{eq:4}
\end{equation}
and
\begin{equation}
 (1-f^{\prime (s)}_{r})= f^{\prime (s)}_{t}\frac{k^{2}_{1}}{k^{2}}\frac{\cos(\tilde\alpha_{t}-\psi)}{\cos(\tilde\alpha-\psi)}
 \label{eq:5}
\end{equation}
 Solving Eqs. \ref{eq:4} and \ref{eq:5} we obtain,

 \begin{equation}
  f^{\prime (s) }_{r}= \frac{k\cos(\tilde\alpha-\psi)-k_{1}\cos(\tilde\alpha_{t}-\psi)}{k\cos(\tilde\alpha-\psi)+k_{1}\cos(\tilde\alpha_{t}-\psi)} \nonumber ,
  \end{equation}
and
\begin{equation}
f^{\prime (s)}_{t}= \left(\frac{k}{k_{1}}\right)^{2}\frac{2k_{1}\cos(\tilde\alpha-\psi)}{k\cos(\tilde\alpha-\psi)+k_{1}\cos(\tilde\alpha_{t}-\psi)}
\end{equation}

Using the above coefficients we now write the reflected electric field expression for each plane wave by adding s and p components of $E_{r}^{\prime\prime\prime}$ as shown in section \ref{eq:flat} . 
\begin{equation}
\vec E^{\prime\prime\prime}_{r}= \vec E^{\prime\prime\prime (s)}_{r}+\vec E^{\prime\prime\prime (p)}_{r} 
\end{equation}
Finally we transform back to the fixed coordinate system  $(x-y-z)$ . Using the inverse of the rotation matrix \textit{Rot}, we write the expression for reflected electric field in the original coordinate system as
\begin{equation}
\vec E_{r}= Rot^{-1} \cdot \vec E^{\prime\prime\prime}_{r}
\end{equation}
Since we are interested only in the perpendicular component, we consider only the y-component of the electric field. For each plane wave we obtain the y- component of $\vec E_{r}$ as 
\begin{equation}\label{eq:ere}
E_{r, y}= 
{ik^3\over 8\epsilon \pi^2} \tilde \Pi_{S,r} \left[f^{\prime (s)}_{r}\cos^{2}\tilde\beta
-f^{\prime (p)}_{r}\cos\tilde\alpha\cos(\tilde\alpha-2\psi)\sin^{2}\tilde\beta
\right]
\end{equation}
We include the corrections due to the roughness of ice surface 
by using the model \cite{ROMEROWOLF2015463,PhysRevD.98.042004} 
\begin{equation}
 F_{rough}(k,\rho,\theta) = \exp\left[{-2k^2\sigma_h(\rho_\perp)^2\cos^2\theta_z}\right] \,.
\label{eq:Frough}
\end{equation}
Here $\theta_z$ is the angle relative to normal at the point of specular
reflection,
 $\rho_\perp^2 = x_\perp^2+y_\perp^2$, $x_\perp,y_\perp$ represent the coordinates with origin at the specular point and
\begin{equation}
\sigma_h(L) = \sigma_h(L_0)\left({L\over L_0}\right)^H\,.
\end{equation}
We choose the parameters $L_{0} = 150$ m, $\sigma_{h}(150m) = 0.041$m and $H =0.65$ which are found to provide reasonable agreement with data for all elevation angles. Including the roughness factor $F_{rough}=F(k,\rho,\theta)$ in our calculation \cite{PhysRevD.98.042004},
 we compute the y-component of total reflected field. 
The resulting total reflected electric field can be written as
\begin{equation}
\begin{aligned}
E_{(r,total),y}= \int_0^{2\pi}
\int_0^{{\pi\over 2}-i\infty}F_{rough} E_{r,y} 
\sin\alpha d\alpha d\beta 
.
\label{eq:totalspherical}
\end{aligned}
\end{equation} 

\begin{figure}[H]
\begin{centering}
\includegraphics[scale=0.62]{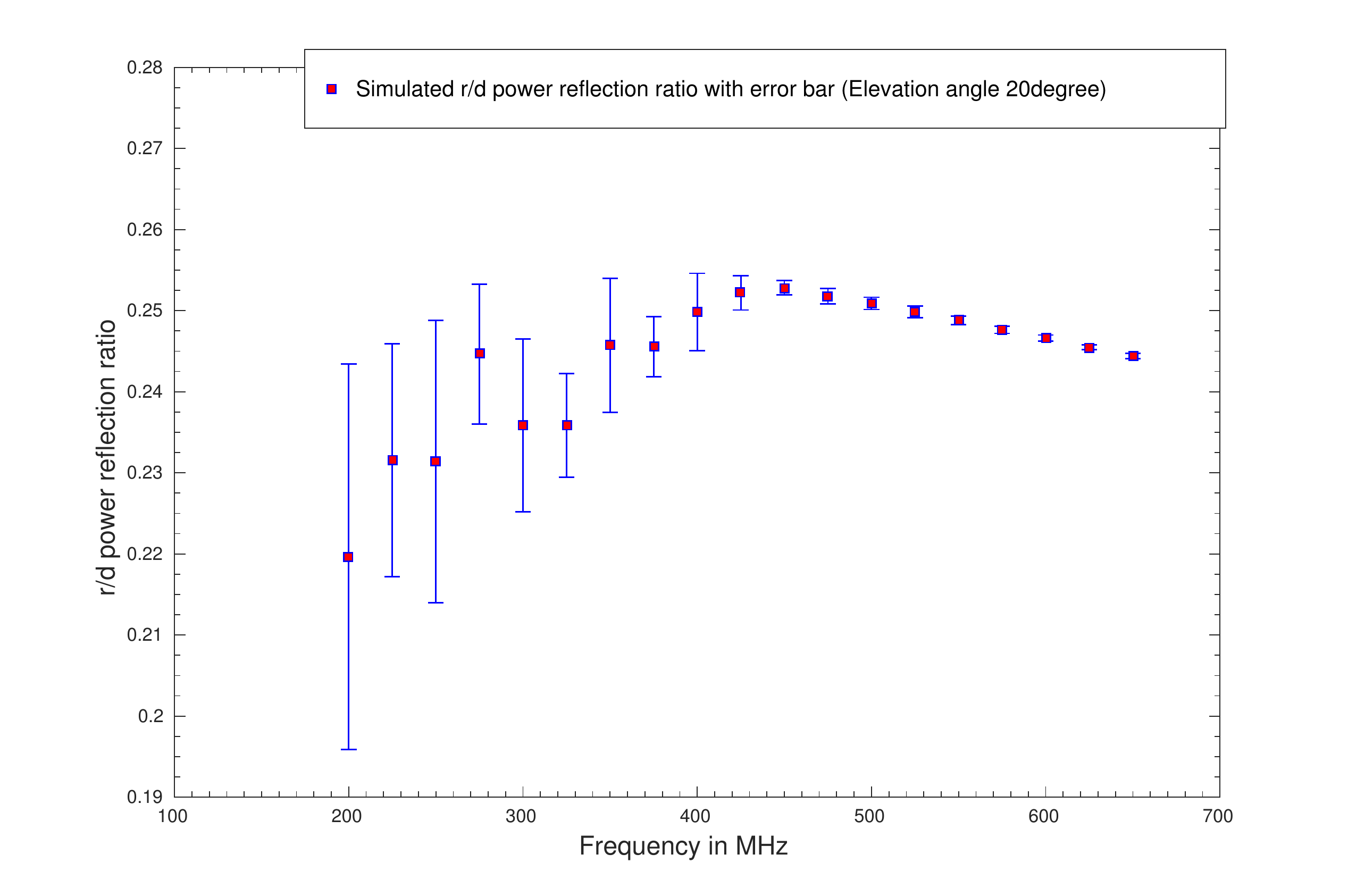}
\par\end{centering}
\caption{ The ratio of reflected to direct power ($r/d$) as a function of frequency for 
n=1.4 and elevation angle (angle with respect to ground) of 20 degrees. }
\label{fig:freqdep}
\end{figure}

The integral gets dominant contribution from regions close to the
specular point. Hence we integrate only over a small neighbourhood of this point.
We choose refractive index of ice $n=1.4$ and compute Eq. \ref{eq:totalspherical} for several frequencies in the range $200$ MHz-$650$ MHz. The power reflection ratio shows a mild dependence
on frequency. This is shown in Fig. \ref{fig:freqdep}. We see that the ratio
 first shows a mild increase with frequency and then starts to decrease. The error bars
shown arise since the reflected power shows rapid fluctuations as a function
of frequency as well as the angle of incidence. 
This was also seen in the formalism used in  \cite{PhysRevD.98.042004} 
for a spherical surface.
The level of fluctuations depend on the roughness parameters and get reduced with increase in the roughness contribution. 
In the earlier formalism \cite{PhysRevD.98.042004}, they were found to be negligible for the roughness parameters 
used in this calculation. This is not the case for the present formalism. 
We also find that the amplitude of the fluctuations becomes very large for
small elevation angles as well as for small frequencies. In Fig. \ref{fig:freqdep} we 
show the ratio of reflected to direct power, averaged over a small neighbourhood of the chosen frequency. The corresponding
standard deviation gives the error.  
We also find that the
 ratio increases slightly with the refractive index of ice for large values of $\theta_{z}$. 

\begin{figure}[H]
\begin{centering}
\includegraphics[scale=0.52]{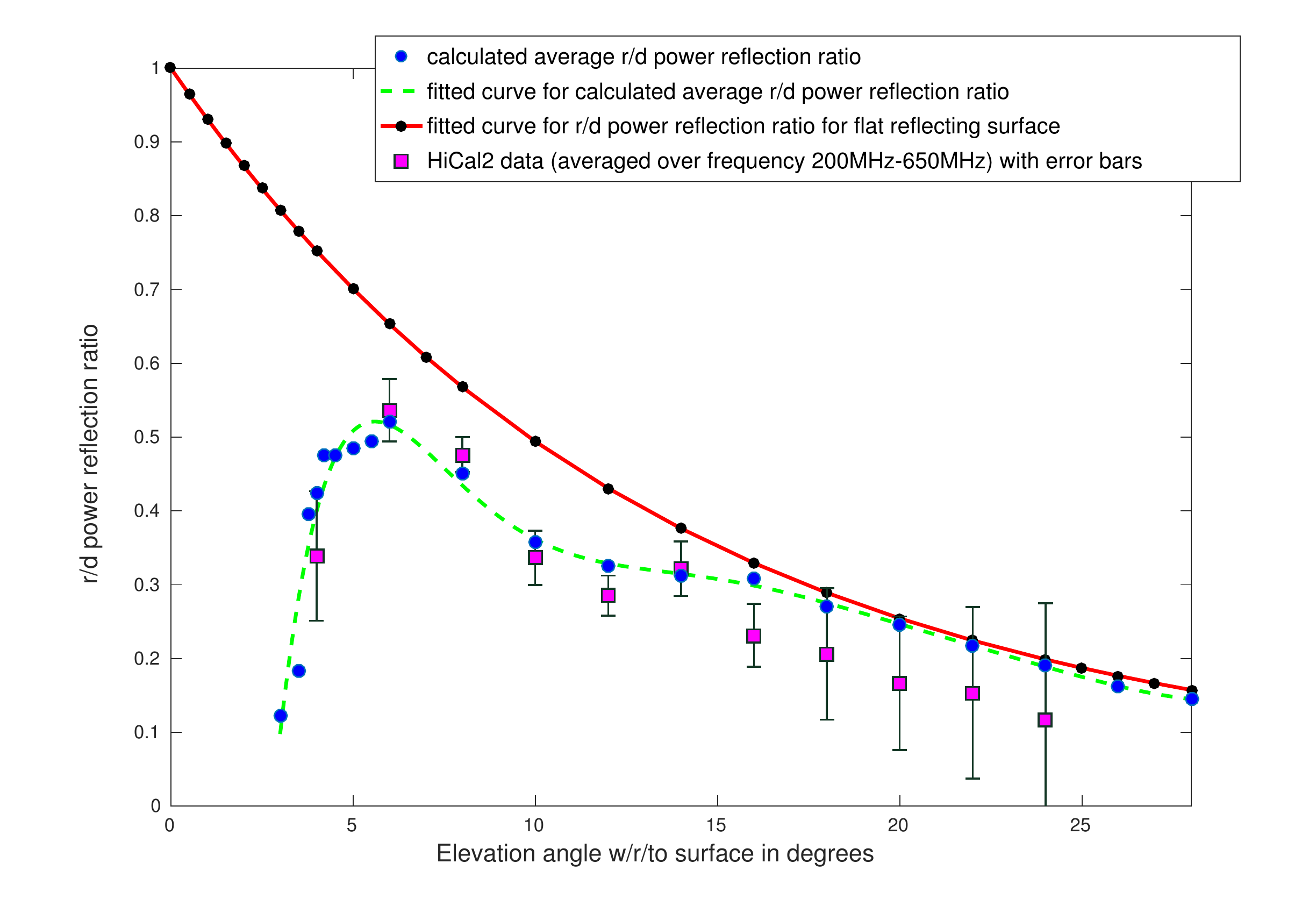}
\par\end{centering}
\caption{ Comparison of the computed (blue dots) average power reflectance 
ratio $r/d$ as a function of elevation angle relative to the ground with n=1.4 and frequency in the range $200-650$ MHz with the HiCal-2 experimental results
(magenta squares with error bars).   The roughness parameters have been chosen to provide a rough match with data without making a detailed fit. The power reflection ratio for a flat surface (without roughness) is shown (red curve and black dots) for comparison. The errors in the theoretical calculation (blue dots)
are given in text.}
\label{fig:spherical_result}
\end{figure}

In Fig. \ref{fig:spherical_result} we show power reflection ratio as a function of the elevation angle $(90^{\rm o}-\theta_z)$. 
This is computed by taking the average of a large number of points (approximately 15) over the
frequency range 200-650 MHz.  
The blue circles show the calculated power ratio and the dashed curve
a smooth fit through the calculated points. Each of these data points have an error
arising from the fluctuations in the theoretical calculation as a function of
frequency as well as elevation angle. The fluctuations are large for small angles
and hence we expect larger error bars in this limit. The error is found to be about
14\% for small elevation angles and reduces to about 4\% for large angles.  
The experimental
data points are from HiCal2 \cite{PhysRevD.98.042004}.  
The result for the flat surface (section \ref{eq:flat}) is shown for comparison.
The only adjustable parameters in our calculation are the parameters in the roughness
model and the refractive index. The refractive index has been set equal to the measured
value of 1.4. The roughness parameters have been taken to be the same as used in earlier
calculations \cite{PhysRevD.98.042004}.  
We see that our theoretical prediction relying on the local plane wave approximation and HiCal-2 data are in good agreement with one another.
In more detail, we find that at large elevation angles the calculated values are systematically
higher than data but agree within errors. The agreement is found to be better at smaller elevation angles. The computed
values depend on the choice of parameters in the roughness model and it is possible to make 
the agreement better by adjusting these parameters. However we do not find much motivation
to do so since the agreement is already very good with the default parameters being used
in the literature.

\section{Implications for ANITA Mystery Events}
So far we have considered monochromatic spherical electromagnetic waves and determined their 
reflection from ice as a function of frequency and the angle of reflection. In the ANITA
observations we need to consider electromagnetic pulses. In this section we consider the
reflection of such pulses from air-ice interface. 
We represent each pulse as a superposition of monochromatic spherical waves, which can
be treated by the procedure described in earlier sections. The Hertz potential for a
monochromatic spherical wave polarized along $\hat y$ is given in Eq. \ref{eq:HertzDirect}. 
The corresponding electric field in the $x-z$ plane in the far zone
can be written as 
\begin{equation}
\vec{E}_{dir}=k^2\frac{e^{ikr}}{4\epsilon\pi r} \hat{y}
\label{eq:Edirect}
\end{equation}

Let $f(p)$ represent the pulse in time domain. The Fourier transform of the pulse can be
represented as
\begin{equation}
\tilde F(n) = \sum_{p=0}^{N-1} f(p) e^{i{2\pi n\over N} p}
\end{equation}
where $N$ represents the total number of samples in the time domain. We decompose the Fourier
transform into the real and imaginary part
\begin{equation}
\tilde F(n) = \tilde F_{real}(n)  + i\tilde F_{imag}(n)
\label{eq:Fouriercomp}
\end{equation}
The direct electric field at the propagation distance
 $r = \sqrt{x^{2}+y^{2}+(z-z_{0})^2}$ 
can be expressed in terms of these by the inverse Fourier transform,
\begin{eqnarray}
E_{d}[p,r] &=& \frac{1}{N}\sum_{n=0}^{N-1} [\tilde{F}_{real}[n]+i\tilde{F}_{imag}[n]]\frac{e^{i(kr-\omega(t+t_{0}))}}{r}\nonumber\\
&=&\frac{1}{Nr}\sum_{n=0}^{N-1} [\tilde{F}_{real}[n]+i\tilde{F}_{imag}[n]]e^{-i\frac{2\pi n p}{N}}
\label{eq:fourier4}
 \end{eqnarray}
where $\omega = kc$ and we have set $t_0 = kr/\omega$ in order to remove an overall phase.  
We have also absorbed the overall factor $k^2/4\pi\epsilon$ in 
Eq. \ref{eq:Edirect} into $\tilde{F}_{real}[n]$ and $\tilde{F}_{imag}[n]$. 

For the reflected pulse, we decompose each monochromatic spherical wave
using Eq. \ref{eq:decomposition} and determine the corresponding 
reflected wave. For each plane wave the $y$-component of the reflected
wave is given by Eq. \ref{eq:ere}. Integrating over all the plane
waves, the total reflected field is given by
\begin{equation}
 E^{\prime}_{ref,y}= \int_{0}^{\frac{\pi}{2}-i\infty}\int_{0}^{2\pi}\int_{\omega}
{ik\over 2\pi} \tilde \Pi_{S,r} F_{rough}(\omega,\alpha,\beta,\theta_{z})\eta(\alpha, \beta, \omega)(\tilde{F}(\omega)e^{-i\omega(t+t_{0})})d\omega d\Omega \,.
\label{eq:eref_1}
\end{equation}
where $d\Omega= \sin\alpha d\alpha d\beta$, $\tilde F(\omega)$ is the continuous Fourier transform and 
\begin{equation}
\eta(\alpha, \beta, \omega) = 
\left[f^{\prime s}_{r}\cos^{2}\tilde\beta
-f^{\prime p}_{r}\cos\tilde\alpha\cos(\tilde\alpha-2\psi)\sin^{2}\tilde\beta)
\right], 
\end{equation}
$\tilde \Pi_{S,r}$ is given in Eq. \ref{eq:Pisr} and 
$F_{rough}$ is given in Eq. \ref{eq:Frough}. 
The final expression for the electric field in terms of the Fourier components (Eq. \ref{eq:Fouriercomp})
is given by 
\begin{eqnarray}
\label{eq:final_eref1}
E_{ref,y}[p]= \frac{1}{N}\sum_{n=0}^{N-1} \frac{k}{2 \pi}\left(i\tilde{F}_{real}[n]-\tilde{F}_{imag}[n]\right)
e^{-i\omega t}\chi(\omega,\alpha,\beta) \\
=\frac{1}{N}\sum_{n=0}^{N-1}\chi(\omega,\alpha,\beta)\frac{k}{2 \pi} 
\left(i\tilde{F}_{real}[n]-\tilde{F}_{imag}[n]\right)  e^{-i\frac{2\pi n p}{N}}
\label{eq:fourier5}
 \end{eqnarray}
where
\begin{equation}
\label{eq:final_eref2}
\chi(\omega,\alpha,\beta)=\int_{0}^{\frac{\pi}{2}-i\infty}\int_{0}^{2\pi}\left[F_{rough}\tilde \Pi_{S,r}
e^{-i\omega t_0}\left[f^{\prime s}_{r}\cos^{2}\tilde\beta
-f^{\prime p}_{r}\cos\tilde\alpha\cos(\tilde\alpha-2\psi)\sin^{2}\tilde\beta)
\right]d \Omega \right]
\end{equation}
We set $t_0= (r_1+r_2)/c$ where $r_1$ and $r_2$ are respectively the distances of the source
and the detector from the specular point. 
From this we extract the real part of the final reflected electric field.
We clarify that the integral over $\alpha$ receives dominant contributions only from a small
region close to the specular point. Hence we do not need to integrate over complex values of $\alpha$.  

\begin{figure}[H]
\begin{centering}
\includegraphics[scale=0.7]{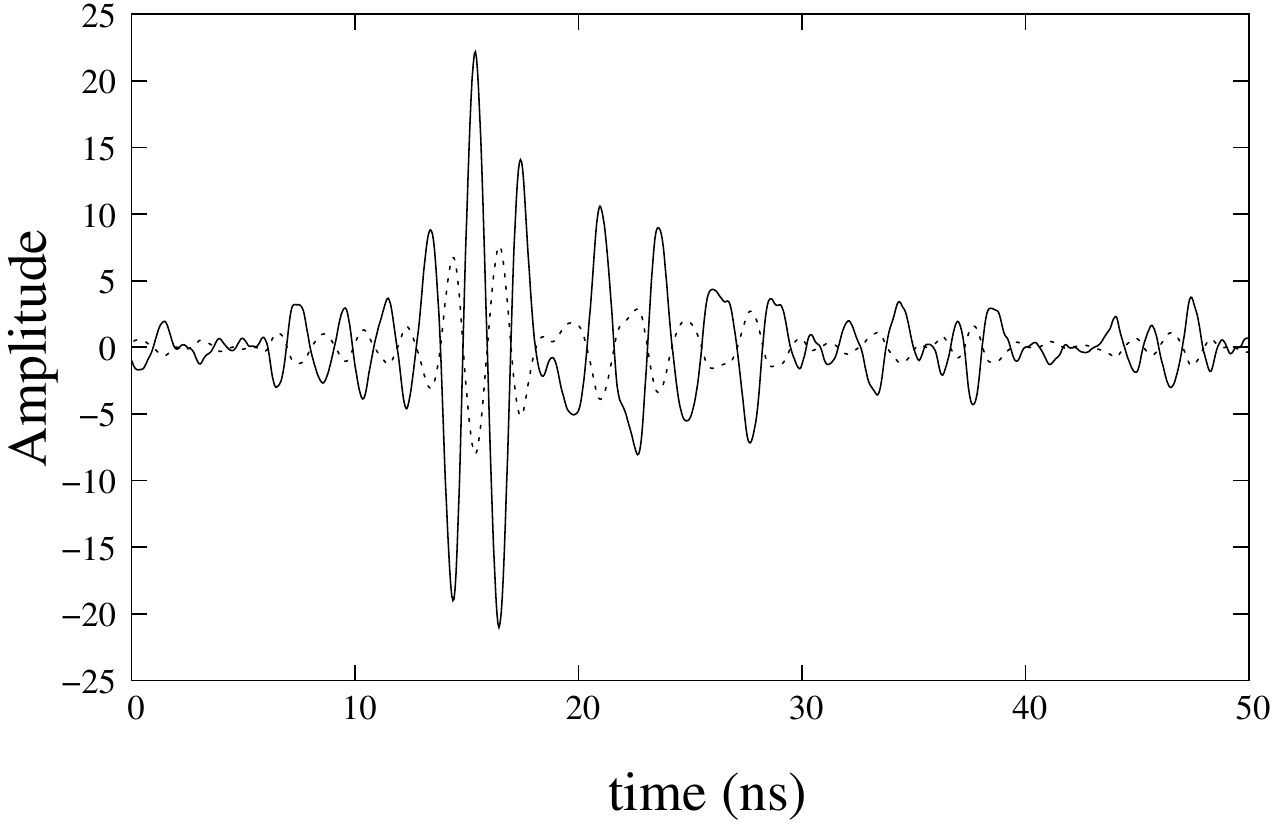}
\par\end{centering}
\caption{The direct pulse (solid line) and the reflected pulse (dashed line)
using the roughness model given in Eq. \ref{eq:Frough} and reflection angle of 78$^{\rm o}$}
\label{fig:isoroughness}
\end{figure}

In Fig. \ref{fig:isoroughness} we show the result for a particular HiCal-1 
pulse assuming a reflection
angle of 78$^{\rm o}$ using the roughness model given in Eq. \ref{eq:Frough}. As expected
we find that the reflected pulse is 180$^{\rm o}$ out of phase with the direct pulse. Although this is to be expected, it is 
not entirely clear whether it will continue to hold in the presence of 
surface roughness effects.  
Hence it is reassuring that the effect emerges in a 
rigourous formalism without making
any uncontrolled approximations and provides further justification for
the claim that the ANITA mystery events \cite{PhysRevLett.117.071101} 
require Physics beyond the Standard Model \cite{Cherry:2018rxj,2018arXiv180311554A,Huang:2018als,Fox:2018syq,Chauhan:2018lnq}. 

\begin{figure}[H]
\begin{centering}
\includegraphics[scale=0.7]{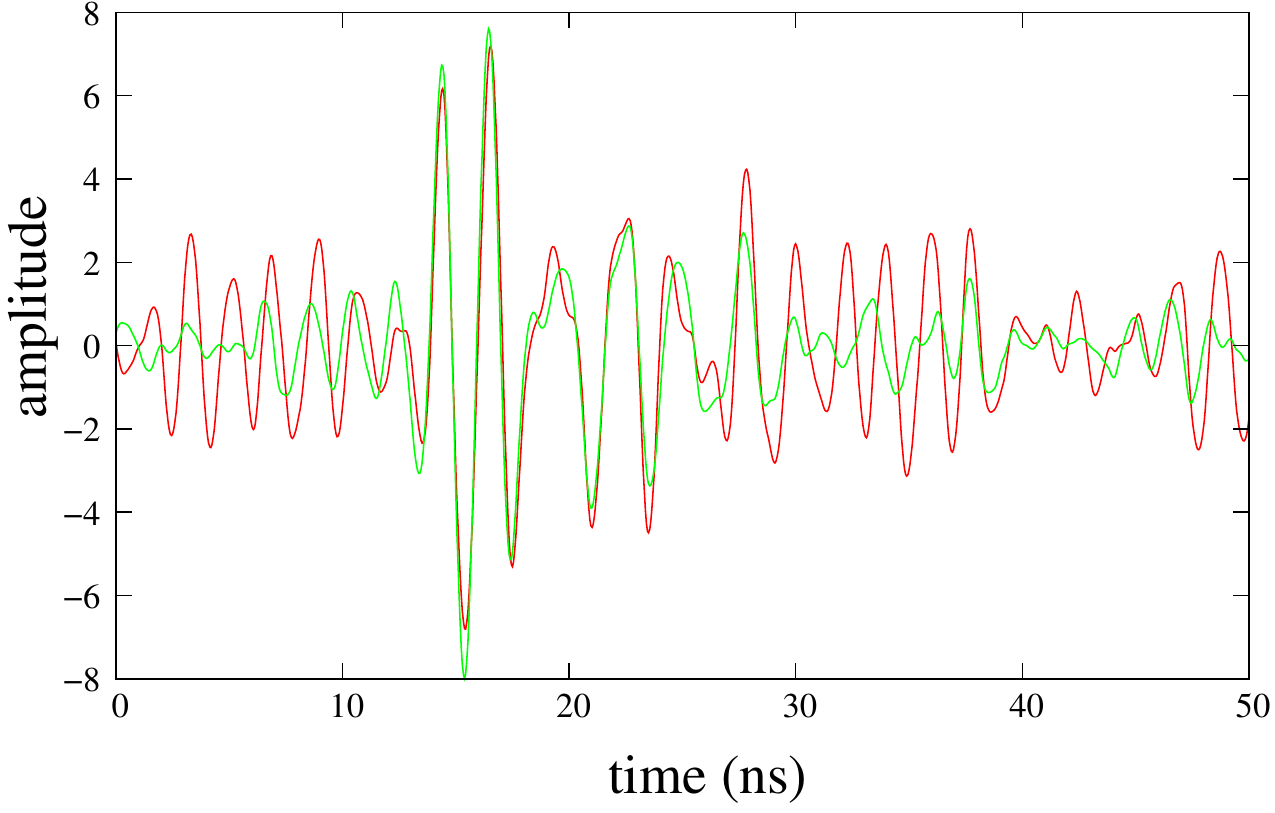}
\par\end{centering}
\caption{The reflected pulse for the case of roughness model in Eq. \ref{eq:Frough} (green line) 
and the model in Eq. \ref{eq:rough2} (red line)
with reflection angle of 78$^{\rm o}$.}
\label{fig:aniso_rough}
\end{figure}

Our formalism allows us to examine more complicated roughness models.
A detailed investigation along this line is beyond the scope of
the present paper. Here we examine   
how the pulse changes if we make the roughness model asymmetric  
by setting
\begin{equation} 
\rho_\perp^2 \rightarrow x_\perp^2 + \xi^2 y_\perp^2 
\label{eq:rough2}
\end{equation} 
in Eq. \ref{eq:Frough}. This model essentially leads to different smoothness in different 
directions. The result for $\xi=0.25$ is shown in Fig. \ref{fig:aniso_rough}. We find that the
main change is the amplitude of the different peaks and dips without any effect on the phase. Hence we find that an asymmetric model of roughness also preserves the
phase inversion of the reflected wave.  
 However we do see a significant distortion in the pulse shape with considerable
power leaking out from the central pulse.
The relative heights of the dominant peak and the dip get inverted as we change $\xi$ from 1 to
0.25. Hence we see that there is some possibility of misidentification of the reflected pulse
as a direct pulse even though the phase is not inverted. However if we view the actual pulses 
observed in ANITA (see Fig. 2 of \cite{Gorham:2018ydl}), we find that the difference in amplitude
between the dominant peak and dip is substantial. 
The roughness effects are relatively small and cannot 
 distort such a 
pulse to the extent that it can be misidentified. We also   
try out two other models with the following replacements
in Eq. \ref{eq:Frough}: 
(i) $\vec\rho_\perp \rightarrow ( \vec\rho_\perp -\vec\rho_{\perp 0})$; 
(ii) $k^2\rightarrow (k-k_0)^2$
where $\vec\rho_{\perp 0}$  and $k_0$ are constants. The conclusions in
both these cases are same as those with model given in Eq. \ref{eq:rough2}.   
In any case it may be worth investigating
this further to see whether we can completely rule out the 
possibility of such  
 a misidentification. We postpone this to future work.

\section{Conclusions}
In this paper we have developed a reliable formalism to handle reflection
of spherical electromagnetic wave from a spherical surface. The treatment
is based on an expansion of the spherical wave in terms of plane waves. 
Each plane wave is reflected from the curved surface by assuming that 
the reflected wave can be approximated as a plane wave in the neighbourhood
of any point. Globally the reflected wave is not a plane wave. The final 
result is obtained by integrating over all the reflected waves corresponding
to different incident plane waves. The procedure involves no uncontrolled
approximation. We apply it to ANITA HiCal-2 observations using a 
reasonable roughness model. We find that our theoretical results for
power reflection ratio
are in good agreement with data for all the elevation angles observed
at HiCal-2. In general we find that our results are close to Fresnel reflection
for 
large elevation angles and, as expected, deviate considerably from 
Fresnel for small angles. The theoretical calculation also shows rapid
oscillations as a function of frequency as well as elevation angle. 
The fluctuations are found to be rather large for small elevation angles.
The final results are obtained after averaging over the frequency 
range 200-650 MHz relevant for HiCal-2 data. 
We also apply our procedure in order to determine the reflected pulse
profile using as template the pulses generated by HiCal. The main aim
of this study is to compare the reflected pulse shape with the incident
pulse and to determine if in some cases a reflected pulse can be misidentified
as direct. This has application to the observed mystery events by ANITA. 
We find that the roughness effects can lead to a significant distortion
of the signal such that the relative strengths of dominant peak and trough
in the reflected wave
can get inverted. However the effect is relatively small and can arise only
in cases in which these two amplitudes in the direct pulse are not too 
different. This does not appear to be applicable to the real cosmic ray
pulses observed by ANITA. Hence we conclude that roughness effects may
not lead to misidentification of ANITA pulses and hence may not provide an
 explanation for the observed mystery events. 
However more work is required in order to completely rule out this possibility.

\section{Acknowledgement}
We thank David Besson and Steven Prohira for valuable inputs and discussions throughout this work. We also thank the entire ANITA collaboration for providing 
us with HiCal-2 reflectivity results and HiCal-1 pulses that we 
use in this paper.

\bibliographystyle{apsrev}
\bibliography{references2}

\begin{thebibliography}{22}
\expandafter\ifx\csname natexlab\endcsname\relax\def\natexlab#1{#1}\fi
\expandafter\ifx\csname bibnamefont\endcsname\relax
  \def\bibnamefont#1{#1}\fi
\expandafter\ifx\csname bibfnamefont\endcsname\relax
  \def\bibfnamefont#1{#1}\fi
\expandafter\ifx\csname citenamefont\endcsname\relax
  \def\citenamefont#1{#1}\fi
\expandafter\ifx\csname url\endcsname\relax
  \def\url#1{\texttt{#1}}\fi
\expandafter\ifx\csname urlprefix\endcsname\relax\def\urlprefix{URL }\fi
\providecommand{\bibinfo}[2]{#2}
\providecommand{\eprint}[2][]{\url{#2}}

\bibitem[{\citenamefont{Gorham et~al.}(2010)}]{Gorham:2010kv}
\bibinfo{author}{\bibfnamefont{P.~W.} \bibnamefont{Gorham}}
  \bibnamefont{et~al.} (\bibinfo{collaboration}{ANITA}),
  \bibinfo{journal}{Phys. Rev.} \textbf{\bibinfo{volume}{D82}},
  \bibinfo{pages}{022004} (\bibinfo{year}{2010}), \bibinfo{note}{[Erratum:
  Phys. Rev. D85, 049901 (2012)]}, \eprint{1003.2961}.

\bibitem[{\citenamefont{Gorham et~al.}(2009)\citenamefont{Gorham, Allison,
  Barwick, Beatty, Besson et~al.}}]{GORHAM200910}
\bibinfo{author}{\bibfnamefont{P.}~\bibnamefont{Gorham}},
  \bibinfo{author}{\bibfnamefont{P.}~\bibnamefont{Allison}},
  \bibinfo{author}{\bibfnamefont{S.}~\bibnamefont{Barwick}},
  \bibinfo{author}{\bibfnamefont{J.}~\bibnamefont{Beatty}},
  \bibinfo{author}{\bibfnamefont{D.}~\bibnamefont{Besson}},
  \bibnamefont{et~al.}, \bibinfo{journal}{Astroparticle Physics}
  \textbf{\bibinfo{volume}{32}}, \bibinfo{pages}{10 } (\bibinfo{year}{2009}),
  ISSN \bibinfo{issn}{0927-6505},
  \urlprefix\url{http://www.sciencedirect.com/science/article/pii/S09276505090%
00838}.

\bibitem[{\citenamefont{Gorham et~al.}(2016)\citenamefont{Gorham, Nam,
  Romero-Wolf, Hoover et~al.}}]{PhysRevLett.117.071101}
\bibinfo{author}{\bibfnamefont{P.~W.} \bibnamefont{Gorham}},
  \bibinfo{author}{\bibfnamefont{J.}~\bibnamefont{Nam}},
  \bibinfo{author}{\bibfnamefont{A.}~\bibnamefont{Romero-Wolf}},
  \bibinfo{author}{\bibfnamefont{S.}~\bibnamefont{Hoover}},
  \bibnamefont{et~al.} (\bibinfo{collaboration}{ANITA Collaboration}),
  \bibinfo{journal}{Phys. Rev. Lett.} \textbf{\bibinfo{volume}{117}},
  \bibinfo{pages}{071101} (\bibinfo{year}{2016}),
  \urlprefix\url{https://link.aps.org/doi/10.1103/PhysRevLett.117.071101}.

\bibitem[{\citenamefont{Gorham et~al.}(2012)\citenamefont{Gorham, Allison,
  Baughman, Beatty et~al.}}]{PhysRevD.85.049901}
\bibinfo{author}{\bibfnamefont{P.~W.} \bibnamefont{Gorham}},
  \bibinfo{author}{\bibfnamefont{P.}~\bibnamefont{Allison}},
  \bibinfo{author}{\bibfnamefont{B.~M.} \bibnamefont{Baughman}},
  \bibinfo{author}{\bibfnamefont{J.~J.} \bibnamefont{Beatty}},
  \bibnamefont{et~al.}, \bibinfo{journal}{Phys. Rev. D}
  \textbf{\bibinfo{volume}{85}}, \bibinfo{pages}{049901}
  (\bibinfo{year}{2012}),
  \urlprefix\url{https://link.aps.org/doi/10.1103/PhysRevD.85.049901}.

\bibitem[{\citenamefont{Hoover et~al.}(2010)\citenamefont{Hoover, Nam, Gorham,
  Grashorn et~al.}}]{PhysRevLett.105.151101}
\bibinfo{author}{\bibfnamefont{S.}~\bibnamefont{Hoover}},
  \bibinfo{author}{\bibfnamefont{J.}~\bibnamefont{Nam}},
  \bibinfo{author}{\bibfnamefont{P.~W.} \bibnamefont{Gorham}},
  \bibinfo{author}{\bibfnamefont{E.}~\bibnamefont{Grashorn}},
  \bibnamefont{et~al.}, \bibinfo{journal}{Phys. Rev. Lett.}
  \textbf{\bibinfo{volume}{105}}, \bibinfo{pages}{151101}
  (\bibinfo{year}{2010}),
  \urlprefix\url{https://link.aps.org/doi/10.1103/PhysRevLett.105.151101}.

\bibitem[{\citenamefont{Askar'yan}(1962)}]{Askaryan:1962hbi}
\bibinfo{author}{\bibfnamefont{G.~A.} \bibnamefont{Askar'yan}},
  \bibinfo{journal}{Sov. Phys. JETP} \textbf{\bibinfo{volume}{14}},
  \bibinfo{pages}{441} (\bibinfo{year}{1962}), \bibinfo{note}{[Zh. Eksp. Teor.
  Fiz.41,616(1961)]}.

\bibitem[{\citenamefont{{Askaryan}}(1962)}]{1962JPSJS..17C.257A}
\bibinfo{author}{\bibfnamefont{G.~A.} \bibnamefont{{Askaryan}}},
  \bibinfo{journal}{Journal of the Physical Society of Japan Supplement}
  \textbf{\bibinfo{volume}{17}}, \bibinfo{pages}{257} (\bibinfo{year}{1962}).

\bibitem[{\citenamefont{{Askar'yan}}(1965)}]{1965JETP...21..658A}
\bibinfo{author}{\bibfnamefont{G.~A.} \bibnamefont{{Askar'yan}}},
  \bibinfo{journal}{Soviet Journal of Experimental and Theoretical Physics}
  \textbf{\bibinfo{volume}{21}}, \bibinfo{pages}{658} (\bibinfo{year}{1965}).

\bibitem[{\citenamefont{Prohira et~al.}(2018)\citenamefont{Prohira, Novikov,
  Dasgupta, Jain, Nande et~al.}}]{PhysRevD.98.042004}
\bibinfo{author}{\bibfnamefont{S.}~\bibnamefont{Prohira}},
  \bibinfo{author}{\bibfnamefont{A.}~\bibnamefont{Novikov}},
  \bibinfo{author}{\bibfnamefont{P.}~\bibnamefont{Dasgupta}},
  \bibinfo{author}{\bibfnamefont{P.}~\bibnamefont{Jain}},
  \bibinfo{author}{\bibfnamefont{S.}~\bibnamefont{Nande}}, \bibnamefont{et~al.}
  (\bibinfo{collaboration}{ANITA Collaboration}), \bibinfo{journal}{Phys. Rev.
  D} \textbf{\bibinfo{volume}{98}}, \bibinfo{pages}{042004}
  (\bibinfo{year}{2018}),
  \urlprefix\url{https://link.aps.org/doi/10.1103/PhysRevD.98.042004}.

\bibitem[{\citenamefont{Besson et~al.}(2015)\citenamefont{Besson, Stockham
  et~al.}}]{doi:10.1002/2013RS005315}
\bibinfo{author}{\bibfnamefont{D.~Z.} \bibnamefont{Besson}},
  \bibinfo{author}{\bibfnamefont{J.}~\bibnamefont{Stockham}},
  \bibnamefont{et~al.}, \bibinfo{journal}{Radio Science}
  \textbf{\bibinfo{volume}{50}}, \bibinfo{pages}{1} (\bibinfo{year}{2015}).

\bibitem[{\citenamefont{Gorham et~al.}(2017)\citenamefont{Gorham, Allison
  et~al.}}]{doi:10.1142/S2251171717400025}
\bibinfo{author}{\bibfnamefont{P.~W.} \bibnamefont{Gorham}},
  \bibinfo{author}{\bibfnamefont{P.}~\bibnamefont{Allison}},
  \bibnamefont{et~al.}, \bibinfo{journal}{Journal of Astronomical
  Instrumentation} \textbf{\bibinfo{volume}{06}}, \bibinfo{pages}{1740002}
  (\bibinfo{year}{2017}).

\bibitem[{\citenamefont{Schoorlemmer et~al.}(2016)\citenamefont{Schoorlemmer,
  Belov et~al.}}]{SCHOORLEMMER201632}
\bibinfo{author}{\bibfnamefont{H.}~\bibnamefont{Schoorlemmer}},
  \bibinfo{author}{\bibfnamefont{K.}~\bibnamefont{Belov}},
  \bibnamefont{et~al.}, \bibinfo{journal}{Astroparticle Physics}
  \textbf{\bibinfo{volume}{77}}, \bibinfo{pages}{32 } (\bibinfo{year}{2016}),
  ISSN \bibinfo{issn}{0927-6505},
  \urlprefix\url{http://www.sciencedirect.com/science/article/pii/S09276505160%
00025}.

\bibitem[{\citenamefont{{Gorham} et~al.}(2017)\citenamefont{{Gorham},
  {Allison}, {Banerjee}, {Batten}, {Beatty}, {Belov}, {Besson}
  et~al.}}]{2017arXiv171011175G}
\bibinfo{author}{\bibfnamefont{P.~W.} \bibnamefont{{Gorham}}},
  \bibinfo{author}{\bibfnamefont{P.}~\bibnamefont{{Allison}}},
  \bibinfo{author}{\bibfnamefont{O.}~\bibnamefont{{Banerjee}}},
  \bibinfo{author}{\bibfnamefont{L.}~\bibnamefont{{Batten}}},
  \bibinfo{author}{\bibfnamefont{J.~J.} \bibnamefont{{Beatty}}},
  \bibinfo{author}{\bibfnamefont{K.}~\bibnamefont{{Belov}}},
  \bibinfo{author}{\bibfnamefont{D.~Z.} \bibnamefont{{Besson}}},
  \bibnamefont{et~al.}, \bibinfo{journal}{ArXiv e-prints}
  (\bibinfo{year}{2017}), \eprint{1710.11175}.

\bibitem[{\citenamefont{Wang}(2018)}]{Wang:2017fnm}
\bibinfo{author}{\bibfnamefont{S.-H.} \bibnamefont{Wang}}
  (\bibinfo{collaboration}{ARIANNA, TAROGE}), \bibinfo{journal}{PoS}
  \textbf{\bibinfo{volume}{ICRC2017}}, \bibinfo{pages}{358}
  (\bibinfo{year}{2018}).

\bibitem[{\citenamefont{Stratton}(1941)}]{Stratton:105763}
\bibinfo{author}{\bibfnamefont{J.~A.} \bibnamefont{Stratton}},
  \emph{\bibinfo{title}{{Electromagnetic theory}}}, International series in
  pure and applied physics (\bibinfo{publisher}{McGraw-Hill},
  \bibinfo{address}{New York, NY}, \bibinfo{year}{1941}),
  \urlprefix\url{https://cds.cern.ch/record/105763}.

\bibitem[{\citenamefont{Romero-Wolf et~al.}(2015)\citenamefont{Romero-Wolf,
  Vance, Maiwald, Heggy, Ries, and Liewer}}]{ROMEROWOLF2015463}
\bibinfo{author}{\bibfnamefont{A.}~\bibnamefont{Romero-Wolf}},
  \bibinfo{author}{\bibfnamefont{S.}~\bibnamefont{Vance}},
  \bibinfo{author}{\bibfnamefont{F.}~\bibnamefont{Maiwald}},
  \bibinfo{author}{\bibfnamefont{E.}~\bibnamefont{Heggy}},
  \bibinfo{author}{\bibfnamefont{P.}~\bibnamefont{Ries}}, \bibnamefont{and}
  \bibinfo{author}{\bibfnamefont{K.}~\bibnamefont{Liewer}},
  \bibinfo{journal}{Icarus} \textbf{\bibinfo{volume}{248}}, \bibinfo{pages}{463
  } (\bibinfo{year}{2015}), ISSN \bibinfo{issn}{0019-1035},
  \urlprefix\url{http://www.sciencedirect.com/science/article/pii/S00191035140%
06009}.

\bibitem[{\citenamefont{Cherry and Shoemaker}(2018)}]{Cherry:2018rxj}
\bibinfo{author}{\bibfnamefont{J.~F.} \bibnamefont{Cherry}} \bibnamefont{and}
  \bibinfo{author}{\bibfnamefont{I.~M.} \bibnamefont{Shoemaker}}
  (\bibinfo{year}{2018}), \eprint{1802.01611}.

\bibitem[{\citenamefont{{Anchordoqui} et~al.}(2018)\citenamefont{{Anchordoqui},
  {Barger}, {Learned}, {Marfatia}, and {Weiler}}}]{2018arXiv180311554A}
\bibinfo{author}{\bibfnamefont{L.~A.} \bibnamefont{{Anchordoqui}}},
  \bibinfo{author}{\bibfnamefont{V.}~\bibnamefont{{Barger}}},
  \bibinfo{author}{\bibfnamefont{J.~G.} \bibnamefont{{Learned}}},
  \bibinfo{author}{\bibfnamefont{D.}~\bibnamefont{{Marfatia}}},
  \bibnamefont{and} \bibinfo{author}{\bibfnamefont{T.~J.}
  \bibnamefont{{Weiler}}}, \bibinfo{journal}{arXiv e-prints}
  \bibinfo{eid}{arXiv:1803.11554} (\bibinfo{year}{2018}), \eprint{1803.11554}.

\bibitem[{\citenamefont{Huang}(2018)}]{Huang:2018als}
\bibinfo{author}{\bibfnamefont{G.-y.} \bibnamefont{Huang}},
  \bibinfo{journal}{Phys. Rev.} \textbf{\bibinfo{volume}{D98}},
  \bibinfo{pages}{043019} (\bibinfo{year}{2018}), \eprint{1804.05362}.

\bibitem[{\citenamefont{Fox et~al.}(2018)\citenamefont{Fox, Sigurdsson,
  Shandera, Mészáros, Murase, Mostafá, and Coutu}}]{Fox:2018syq}
\bibinfo{author}{\bibfnamefont{D.~B.} \bibnamefont{Fox}},
  \bibinfo{author}{\bibfnamefont{S.}~\bibnamefont{Sigurdsson}},
  \bibinfo{author}{\bibfnamefont{S.}~\bibnamefont{Shandera}},
  \bibinfo{author}{\bibfnamefont{P.}~\bibnamefont{Mészáros}},
  \bibinfo{author}{\bibfnamefont{K.}~\bibnamefont{Murase}},
  \bibinfo{author}{\bibfnamefont{M.}~\bibnamefont{Mostafá}}, \bibnamefont{and}
  \bibinfo{author}{\bibfnamefont{S.}~\bibnamefont{Coutu}},
  \bibinfo{journal}{Submitted to: Phys. Rev. D}  (\bibinfo{year}{2018}),
  \eprint{1809.09615}.

\bibitem[{\citenamefont{Chauhan and Mohanty}(2018)}]{Chauhan:2018lnq}
\bibinfo{author}{\bibfnamefont{B.}~\bibnamefont{Chauhan}} \bibnamefont{and}
  \bibinfo{author}{\bibfnamefont{S.}~\bibnamefont{Mohanty}}
  (\bibinfo{year}{2018}), \eprint{1812.00919}.

\bibitem[{\citenamefont{Gorham et~al.}(2018)}]{Gorham:2018ydl}
\bibinfo{author}{\bibfnamefont{P.~W.} \bibnamefont{Gorham}}
  \bibnamefont{et~al.} (\bibinfo{collaboration}{ANITA}),
  \bibinfo{journal}{Phys. Rev. Lett.} \textbf{\bibinfo{volume}{121}},
  \bibinfo{pages}{161102} (\bibinfo{year}{2018}), \eprint{1803.05088}.

\end{thebibliography}
 

\end{document}